\documentclass{emulateapj}

\usepackage{amsmath,amssymb,gensymb,times,graphics,morefloats,color}
\usepackage{afterpage, bm}

\usepackage{natbib,hyperref}
\shorttitle{M dwarfs in the NIR}
\shortauthors{Elisabeth R. Newton}

\def\kepler{\emph{Kepler}}
\def\phoenix{PHOENIX} 
\def\spex{SpeX}

\def\Rap{R_{ap}^2}
\def\rmse{\mathrm{RMSE}}

\def\V{$V$}
\def\H{$H$}
\def\K{$K$}
\def\J{$J$}
\def\Z{$Y$}
\def\I{$Z$}

\def\msun{M_{\odot}}

\def\rsun{R_{\odot}}

\def\Vs{V_{tot}}

\def\feh{\mathrm{[Fe/H]}}
\def\dex{\mathrm{\ dex}}
\def\kms{\mathrm{\ km}\mathrm{\ s}^{-1}}
\def\pc{\mathrm{pc}}
\def\dms{\mathrm{D}_\mathrm{MS}}

\def\hind{\mathrm{H}_2\mathrm{O\mbox{-}K2}}
\def\water{\mathrm{H}_2\mathrm{O}}
\def\ewna{\mathrm{EW}_\mathrm{Na}}
\def\na{\mathrm{Na}}

\def\nahind{\mathrm{\ewna}/\left(\hind\right)}
\def\cahind{\mathrm{\mathrm{EW}_\mathrm{Ca}}/\left(\hind\right)}
\def\naco{\mathrm{\ewna/\mathrm{EW}_\mathrm{CO}}}

\def\co{\mathrm{CO}}

\def\la{\langle}
\def\ra{\rangle}

\def\santos{Santos+}
\def\sousa{Sousa+}
\def\spocs{VF05}

\def\nms{447\ }
\def\ncals{36\ }
\def\nlines{27\ }

\begin{document} 
\title{Near-infrared Metallicities, Radial Velocities and Spectral Types for \nms Nearby M Dwarfs}

\author{Elisabeth~R.~Newton\altaffilmark{1}\altaffilmark{*}, David Charbonneau\altaffilmark{1}, Jonathan Irwin\altaffilmark{1}, Zachory K. Berta-Thompson\altaffilmark{1}, Barbara Rojas-Ayala\altaffilmark{2}, Kevin Covey\altaffilmark{3}, James P. Lloyd\altaffilmark{4}}
\altaffiltext{1}{Harvard-Smithsonian Center for Astrophysics, 60 Garden Street, Cambridge, MA 02138, USA}
\altaffiltext{2}{Centro de Astrof'sica, Universidade do Porto, Rua das Estrelas, 4150-762 Porto, Portugal}
\altaffiltext{3}{Lowell Observatory, 1400 West Mars Hill Road, Flagstaff, AZ 86001, USA}
\altaffiltext{4}{Department of Astronomy, Cornell University, 226 Space Sciences Building, Ithaca, NY 14853, USA}
\altaffiltext{*}{\texttt{enewton@cfa.harvard.edu}}

\begin{abstract}
We present metallicities, radial velocities and near-infrared spectral types for \nms M dwarfs determined from moderate resolution ($R\approx2000$) near-infrared (NIR) spectra obtained with IRTF/\spex. These M dwarfs are primarily targets of the MEarth Survey, a transiting planet survey searching for super Earths around mid-to-late M dwarfs within 33pc. We present NIR spectral types for each star and new  spectral templates for IRTF in the \Z, \J, \H\ and \K-bands, created using M dwarfs with near-solar metallicities. We developed two spectroscopic distance calibrations that use NIR spectral type or an index based on the curvature of the \K-band continuum. Our distance calibration has a scatter of 14\%. We searched \nlines NIR spectral lines and 10 spectral indices for metallicity sensitive features, taking into account correlated noise in our estimates of the errors on these parameters. We calibrated our relation using \ncals M dwarfs in common proper pairs with an F, G or K-type star of known metallicity. We validated the physical association of these pairs using proper motions, radial velocities and spectroscopic distance estimates. Our resulting metallicity calibration uses the sodium doublet at $2.2\micron$ as the sole indicator for metallicity. It has an accuracy of $0.12\dex$ inferred from the scatter between the metallicities of the primaries and the estimated metallicities of the secondaries. Our relation is valid for NIR spectral types from M1V to M5V and for $-1.0<\feh<+0.35\dex$. We present a new color-color metallicity relation using $J-H$ and $J-K$ colors that directly relates two observables: the distance from the M dwarf main sequence and equivalent width of the sodium line at $2.2\micron$. We measured radial velocities by modeling telluric features to determine the absolute wavelength calibration of our spectra, and used M dwarf binaries, observations at different epochs, and comparison to precisely measured radial velocities to demonstrate $4\kms$ accuracy.
\end{abstract}


\section{Introduction}

MEarth is a transiting planet survey looking for super Earths around nearby mid to late M dwarfs. As part of our efforts to characterize the local M dwarf population, the MEarth team and collaborators are gathering a diverse data set on these low mass stars. 
These unique data have already begun to bear fruit. \citet{Charbonneau2009} reported the discovery of a super Earth transiting the mid M dwarf GJ 1214. 
\citet[]{Irwin2011} took advantage of our long-baseline photometry to measure rotation periods as long as $120$ days for 41 M dwarfs and investigated their angular momentum evolution, finding that strong winds may be needed to explain the population of slowly rotating field M dwarfs. \citet{Irwin2011a} presented a long period M dwarf-M dwarf eclipsing binary and measured the masses of the two components and the sum of their radii. They find the radii to be inflated by 4\% relative to theoretical predictions, reflecting a well-known problem with stellar models at the bottom of the main sequence \citep[e.g.][]{LopezMorales2007, Boyajian2012}.

Interest in M dwarfs is fueled by prospects for testing theories of planet formation. Creating a planetary system around a small star is one of the simplest ways to test the effect of initial conditions: the disk out of which planets form is less massive around an M dwarf than around a more massive star. Core accretion and gravitational instability models predict different rates of occurrence of planets around low-mass stars, with the formation of giant planets through core accretion being hampered by the low disk surface density and long orbital time scale in M dwarf protoplanetary disks \citep[]{Laughlin2004}. Recent results from \kepler\ showed that giant planets are less likely to be found around K and early M stars than around F and G stars, lending support to the core accretion model \citep[]{Borucki2011,Fressin2013}. A similar finding was reported for M dwarfs targeted by radial velocity surveys \citep[][]{Johnson2007, Cumming2008}. The high metallicity of solar-type stars that host close-in giant planets was confirmed over a decade \citep[e.g.][]{Fischer2005}, but smaller planets have been found around stars of a range of metallicities \citep[]{Buchhave2012}. Efforts have been made to extend these relations to the lowest stellar masses \citep[e.g.][]{Johnson2009,Schlaufman2010,Rojas-Ayala2012}, but have been limited by the small number of planets currently known around M dwarfs.  

M dwarfs present a unique opportunity for the detection and characterization of habitable Earth-sized planets. Mid to late M dwarfs are favorable targets for transiting planet searches  \citep{Nutzman2008}. Their low luminosity puts the habitable zone at smaller orbital radii, making transits more likely and more frequent: for an M4 dwarf, the period of a habitable planet is two weeks, compared to one year for a solar-type star. Because the transit depth is set by the planet-to-star radius ratio, smaller planets are more readily detectable around these stars. The small radius of an M dwarf is also favorable for follow-up studies of an orbiting planet's atmosphere with transmission or occultation techniques and nearby mid M dwarfs are bright enough in the NIR for precise spectroscopic studies \citep[e.g.][]{Bean2011, Crossfield2011, Berta2012}.

In contrast to solar type stars, the physical parameters of M dwarfs are not in general well understood and present a major hurdle for studying transiting planets orbiting M dwarfs. Few M dwarfs are bright enough for direct measurement of their radii \citep[e.g.][]{Berger2006, Boyajian2012}, and discrepancies between the observed radii and theoretical predictions persist \citep[see][for a review]{TorresG.2013}. Empirical calibrations provide an inroad. For example, \citet[]{Muirhead2012a} and \citet[]{Muirhead2012} exploited the \K-band metallicity and temperature relations of \citet[][hereafter R12]{Rojas-Ayala2012} to estimate new planet properties for the \kepler\ Objects of Interest (KOIs) orbiting the coolest \kepler\ stars and discovered the planetary system with the smallest planets currently known, the \kepler-42 system (n\'{e}e KOI-961). \citet[]{Johnson2012} combined existing photometric relations to estimate the stellar properties of KOI-254, one of the few M dwarfs known to host a hot Jupiter. \citet{Ballard2013} used M dwarfs with interferometric radii as a proxy to constrain the radius and effective temperature of \kepler-61b.

Several studies have used M dwarf model atmospheres matched to high resolution spectra to determine stellar parameters. \citet{Woolf2005} estimated M dwarf temperatures and surface gravities from photometry, then, fixing these parameters, inferred the metallicity from the equivalent widths (EWs) of metal lines. Updating and modifying the spectral synthesis method of \citet[]{Valenti1998}, \citet{Bean2006} used TiO and atomic lines in combination with NextGen \phoenix\ model atmospheres \citep{Hauschildt1999} to measure the physical properties of M dwarfs. Most recently, \citet{Onehag2012} matched model spectra from MARCS \citep[]{Gustafsson2008} to observations of FeH molecular features in the infrared and found metallicities higher than those inferred by \citet[]{Bean2006b}. The MARCS model atmospheres do not include dust formation and are not applicable to M dwarfs later than M6V. However, uncertain sources of opacity in the model atmospheres complicate direct interpretation of observed spectra throughout the M spectral class.

An effective technique for quantitatively studying the metallicities of M dwarfs makes use of cool stars in common proper motion (CPM) pairs with an F, G or K-type star, where the primary has a measured metallicity. Assuming the two are coeval, one can infer the metallicity of the low-mass companion and subsequently use a sample of CPM pairs to confirm or empirically calibrate tracers of M dwarf metallicity. \citet[]{Gizis1997} applied this idea to the M subdwarf population, using observations of late-type companions to F and G subdwarfs of known metallicity to confirm the metallicity relation of \citet[]{Gizis1997a}, which used optical spectral indices to infer the metallicity of M subdwarfs.

\citet[]{Bonfils2005} pioneered the empirical calibration of M dwarf metallicities using CPM pairs. The authors found that a metal-rich M dwarf has a redder $V-K$ color at a given absolute \K\ magnitude, due to increased line blanketing by molecular species, particularly TiO and VO. The calibration is valid for $4 < \mathrm{M}_K < 7.5$, $2.5 < V-K < 6$ and $-1.5 < \feh < +0.2 \dex$. \citet[]{Bonfils2005} reported a standard deviation of $0.2 \dex$. \citet[]{Johnson2009}, finding the calibration of \citet{Bonfils2005} to systematically underestimate the metallicities of metal-rich stars, updated the relation by considering the offset from the mean main sequence (MS), assuming the mean MS defined an isometallicity contour with $\feh=-0.05\dex$. Their calibration sample used six metal-rich calibrators. \citet[]{Schlaufman2010} found that the previous works had systematic errors at low and high metallicities and further updated the photometric relation. They used a larger calibration sample comprised only of M dwarfs with precise \V\ magnitudes in CPM pairs with an F, G or K-star, where the primary's metallicity had been determined from high resolution spectroscopy. They also updated the determination of the mean MS, finding that it corresponded to an isometallicity contour with $\feh= -0.14\dex$. However, external information was still required to determine the mean MS. The standard deviation of their fit was $0.15 \dex$.

\citet[]{Neves2012} tested the photometric calibrations of \citet[]{Bonfils2005}, \citet[]{Johnson2009}, and \citet[]{Schlaufman2010} on a new sample of FGK-M CPM pairs that had precise \V-band photometry. With their sample of $23$ M dwarfs, they found the \citet[]{Schlaufman2010} calibration had the lowest residual mean square error ($\rmse=0.19\pm0.03\dex$) and highest correlation coefficient ($\Rap=0.41\pm0.29$), performing marginally better than the \citet{Bonfils2005} calibration. They updated the \citet[]{Schlaufman2010} calibration, though the diagnostic values did not improve by more than the associated errors.

\citet[][hereafter R10]{Rojas-Ayala2010} took a different approach and used moderate resolution \K-band spectra ($R\approx\Delta\lambda/\lambda\approx2700$) to measure metallicity.  They used the EWs of the \ion{Na}{1} doublet and \ion{Ca}{1} triplet to measure metallicity and the $\hind$ index to account for the effects of temperature. The calibration was updated in \citet[][hereafter R12]{Rojas-Ayala2012}, who demonstrated that their empirical metallicities gave reasonable results for solar neighborhood M dwarfs. With $18$ calibrators, this method yielded $\rmse=0.14 \dex$ and $\Rap=0.67$ for their $\feh$ calibration. The lines used in this calibration are isolated across the entire M dwarf spectral sequence and are located near the peak of the M dwarf spectral energy distribution (SED). Parallaxes and accurate magnitudes, which are scarce for M dwarfs, are not required, placing metallicities within reach for many M dwarfs.

\citet[]{Terrien2012} applied the methods of R10 to spectra obtained with the \spex\ instrument on the NASA Infrared Telescope Facility (IRTF), using $22$ CPM pairs as calibrators. They updated the \K-band R10 calibration ($\rmse=0.14\dex$, $\Rap=0.74$) and presented an \H-band calibration ($\rmse=0.14\dex$, $\Rap=0.73$). \citet[]{Mann2013} expanded the sample of calibrators and identified over 100 metal-sensitive features in the NIR and optical. Their calibration sample included $112$ FGK-M CPM pairs, selected on the basis of common proper motion and galactic models. They constructed metallicity relations in the optical and in each of the NIR bands out of metallicity sensitive features and a single parameter to account for temperature dependencies. Their $\feh$ calibrations had standard deviations between $0.11\dex$ and $0.16\dex$ and $\Rap$ values ranging from $0.68$ to $0.86$. They also updated the color-color relation of \citet[]{Johnson2012} and the \K- and \H-band spectroscopic relations of \citet[]{Terrien2012} and R12.

We also note the larger context in which constraints on the physical properties of M dwarfs are applicable. 
For example, \citet[]{Bochanski2007} used SDSS M dwarfs to test the Besan\c{c}on galactic model \citep[]{Robin2003}, comparing observed kinematics to the model and comparing the observed metallicities and active fractions of the thin and thick disk.  In this study and others using SDSS, optical molecular indices were used as a proxy for metallicity \citep[e.g.][]{Gizis1997,Woolf2006}. The $\zeta$-index, which uses CaH and TiO molecular band heads, is commonly used to identify subdwarfs and extreme subdwarfs \citep[]{Lepine2007,Dhital2012}. Theories of star formation must also match the observed luminosity and mass functions of M dwarfs, which are in turn important input into galactic models. \citet[]{Bochanski2010}, again exploiting SDSS, measured the M dwarf luminosity and mass functions. Photometric distance estimates were used in this work, and one of the primary factors complicating these estimates was uncertainty in how metallicity affects absolute magnitude.

In this work, we present our observation and analysis of near infrared (NIR) moderate resolution ($R\approx2000$) spectra of \nms MEarth M dwarfs. Our sample is presented in \S\ref{Sec:sample} and in \S\ref{Sec:spex} we discuss our observations and data reduction. We account for correlated noise when estimating the error on our measurements, as we discuss in \S\ref{Sec:errors}. In \S\ref{Sec:sptypes}, we present by-eye NIR spectral types for each star and a new spectroscopic distance calibration. Our metallicity measurements, described in \S\ref{Sec:metallicity}, are based on the method developed by R12: we use EWs of spectral features in the NIR as empirical tracers of metallicity, using M dwarfs in CPM pairs to calibrate our relationship. We present a color-color metallicity calibration in \S\ref{Sec:colors}. In \S\ref{Sec:rv}, we discuss our method for measuring radial velocities, which uses telluric features to provide the wavelength calibration, and demonstrate $4\kms$ accuracy. Our data are presented in Table A1 and we include updated parameters for those stars observed by R12 in Table A2. We include radial velocities, spectral types and parallaxes compiled from the literature.


\section{Sample}\label{Sec:sample}

Our sample consists of \nms M dwarfs targeted by the MEarth transiting planet survey and 46 M dwarfs in CPM pairs with an F, G or K star of known metallicity, a subset of which we used to calibrate our empirical metallicity relation.

\subsection{MEarth M dwarfs}\label{Sec:MEarth}

The MEarth project is photometrically monitoring 2000 of the nearest mid to late M dwarfs in the northern sky with the goal of finding transiting super Earths. \citet[]{Nutzman2008} described how the MEarth targets were selected from the L\'{e}pine-Shara Proper Motion catalog of northern stars \citep[LSPM-North;][]{Lepine2005a}. For completeness, we summarize their method here. From the subset of stars believed to be within $33$ pc \citep{Lepine2005}, using spectroscopic or photometric distance estimates where parallaxes were unavailable, they selected those with $V-J > 2.3$, $J-K_S > 0.7$, and $J-H > 0.15$, resulting in a sample of probable nearby M dwarfs. The radius for each probable M dwarf was estimated by first using the absolute $K_S$ magnitude-to-mass relation of \citet[]{DelfosseX.2000}, and inputting this mass into the mass-to-radius relationship from \citet[]{Bayless2006}. They subsequently selected all objects with estimated radii below $0.33\rsun$, driven by the desire to maintain sensitivity to planets with radii equal to twice Earth's.

MEarth is a targeted survey, visiting each object with a cadence of 20-30 minutes on each night over one or more observing seasons. A fraction of the sample has sufficient coverage and quality to estimate their rotation periods, with recovered periods ranging from 0.1 to 90 days. These will be discussed in a subsequent paper.

\subsection{Spectroscopy targets}\label{Sec:spectroscopy}

We targeted a subset of the MEarth M dwarfs for NIR spectroscopy. We re-observed 30 stars in common with R12, who focused their efforts on M dwarfs within 8pc, in order to evaluate any systematic differences between our instruments and methods. The IRTF declination limit prevented us from observing stars above $+70\degree$. We divide our targets into four subsamples based on the reason for their selection:
\begin{itemize}
\item{Rotation sample: 181 M dwarfs with preliminary rotation periods measured from MEarth photometry. These show periodic photometric modulation presumed to be due to star spots rotating in and out of view.}
\item{Nearby sample: 257 M dwarfs drawn from the full MEarth sample, for which no clear periodic photometric modulation was detected at the time of selection. This included 131 M dwarfs selected because they have parallaxes available from the literature, 94 M dwarfs with photometric distance estimates, and 32 ``photometrically quiet'' M dwarfs. The photometrically quiet M dwarfs are those for which phase coverage and photometric noise were sufficient to achieve good sensitivity to rotationally induced photometric modulations, but for which no such modulations were observed.}
\item{Metallicity calibrators: 46 M dwarfs in CPM pairs with an F, G or K, where a metallicity measurement is available for the primary. These are discussed in \S\ref{Sec:metallicity}. We used 36 M dwarfs in our final metallicity calibration.} 
\item{Potential calibrators: 10 potential calibrators are in CPM pairs with an F, G or K star but do not have a metallicity measurement available for the primary. We did not include these stars in our metallicity calibration.}
\end{itemize}

We present new observations of $\nms$ nearby M dwarfs in Table A1 (the rotation and nearby samples and potential calibrators). Data for our 46 M dwarf metallicity calibrators are presented separately.


\section{Observations}\label{Sec:spex}

We conducted our observations with the \spex\ instrument on the NASA Infrared Telescope Facility \citep[IRTF]{Rayner2003}. We used the short cross dispersed (SXD) mode with the $0.3\times15\arcsec$ slit. This yielded spectra with $R\approx2000$ covering $0.8-2.4\micron$, with gaps between orders where there is strong atmospheric absorption. Our observations spanned 25 partial nights over 4 semesters. Observing conditions are summarized in Table \ref{Tab:conditions}; in moderate clouds, we observed bright targets. 

We typically acquired four observations of each object, with two observations at each of two nod positions (A and B), in the sequence ABBA. We used the default A position and nod distance, with the A and B positions falling $3\farcs75$ from the edge of the slit (a $7\farcs5$ separation). Most of our targets were observed within half an hour of meridian crossing. For hour angles greater than one, we aligned the slit with the parallactic angle. We observed A0V stars for use as telluric standards within one hour of each science target, at angular separations no more than $15\degree$, and with airmass differences of no more than 0.1 when possible (see \S\ref{Sec:errors}). 
We took flat field spectra (using an internal quartz lamp) and wavelength calibrations (using internal Thorium-Argon lamps) throughout the night, at one hour intervals or after large slews. The typical observation time for a $K=9$ target at each nod was $100$ seconds (for a total integration time of $400$ seconds). Combining four nods yielded a total signal-to-noise ratio (S/N) of 250 per resolution element.

\begin{deluxetable}{l | l l l}
\tablecaption{Observing conditions\label{Tab:conditions}}
\tablecolumns{4}
\tablehead{\colhead{Semester} & \colhead{Start date}  & \colhead{Seeing} & \colhead{Weather conditions}\\
& \colhead{(UT)} & &} 
\startdata
2011A	&May 15	& 	$0.6-1\arcsec$	&	Mostly clear, humid		\\
	&	May 16	&	$0.4-0\farcs8$	&	Some cirrus, humid 	\\
	&	May 17	&	$0\farcs5$		&	Heavy clouds, then clear \\
	&	May 18	&	$0\farcs5$		&	Clear \\
	\hline
2011B	&June 9	&	$0\farcs7$		&	Some clouds \\
	&	Aug 11	&	$0\farcs5$		& 	Some clouds \\
	&	Aug 12	&	$0\farcs5$		&	Heavy clouds  \\
	&	Aug 13	&	$0\farcs5$		&	Mostly clear \\
	&	Aug 14	&	$0\farcs4$		&	Mostly cloudy \\
	&	Oct 7	&	$0\farcs8$		&	Some clouds \\
	&	Oct 8	&	$0\farcs8$		&	Heavy intermittent clouds \\
	&	Oct 9	&	$0\farcs6$		&	Mostly clear \\
	\hline
2012A	&	Feb. 14 &	$1\arcsec$		&	Clear \\
	&	Feb 15 	&	$0.5-1\arcsec$	&	Clear \\
	&	Feb 16 	& 	$0\farcs8$		&	Clear \\
	&	Feb 24	&	$0\farcs8$		& 	Clear \\
	&	Feb 27	&	$1\arcsec$		& 	Heavy intermittent clouds \\
	&	Feb 28	&	$0\farcs8$		&	Clear \\
	&	May 1	&	$0.3-1\farcs2$	&	Clear \\
	&	May 2	&	$0\farcs6$		&	Clear \\
	\hline
2012B	&Aug 14	&	$1-2\arcsec$		&	Clear \\
	&	Aug 26	&	$0\farcs5$		&	Clear \\
	&	Aug 27	&	$0\farcs5$		& 	Clear \\
	&	Jan 26	&	$0\farcs8$		&	Clear	\\
	&	Jan 27	&	$1\farcs1$		&	Heavy morning clouds
\enddata
\end{deluxetable}

We reduced the data with the instrument-specific pipeline \texttt{Spextool} \citep[]{Cushing2004}, modified to allow greater automation and to use higher S/N flat fields, created by median combining all flat field frames from a given night. Images were first flat-field corrected using the master flat from the given night. After subtracting the A and B images, we used boxcar extraction with an aperture radius equal to the full width at half maximum (FWHM) of the average spatial profile and subtracted the residual sky background. To determine the background sky level in the AB subtracted image, we used a linear fit to the regions beginning $1\farcs2$ from the edges of the aperture. This step was important near sunrise and sunset and increasingly important in bluer orders, but the \K-band was largely unaffected. Each spectrum was wavelength calibrated using the set of Thorium-Argon exposure most closely matching in time.

We combined individual spectra for the same object (typically 4 per object) using the \texttt{Spextool} routine \texttt{xcombspec}. We scaled the raw spectra to the median flux level within a fixed wavelength region and removed low order variations in the spectral shapes. We used the highest S/N region of the \H-band for scaling. The modified spectra were combined using the robust weighted mean algorithm, which removed outliers beyond $8\sigma$.

We used \texttt{xtellcor} to perform the telluric corrections \citep[]{Vacca2003}. We used the Paschen $\delta$ line near $1\micron$ in the A0V telluric standard to create a function to describe the instrumental profile and the rotational broadening observed in spectrum. We used \texttt{xtellcor} to convolve this function with a model of Vega and shifted the model to match the star's observed radial velocity. We scaled the line strengths of individual lines to match those observed; for data taken in 2012, we adjusted the scaling by hand. We found this to be a necessary step because even for sub-1\% matches to the Vega model, residual hydrogen lines were apparent. The atmospheric absorption spectrum, as observed by the instrument, was found by dividing the observed A0V spectrum by the modified Vega spectrum. We shifted the atmospheric absorption spectrum to match the absorption features in the object spectrum and divided to remove the atmospheric absorption features present. We performed this step separately in each order, using a region dominated by telluric features to shift the spectra.

We performed flux calibration as part of the telluric correction, but variable weather conditions and slit losses made the absolute flux level unreliable. We do not require absolute flux calibration for our project goals.

\section{Estimation of uncertainty}\label{Sec:errors}

Given the high S/N (typically $>200$) of our spectra, the uncertainties in quantities measured from our data are dominated by correlated noise, rather than random photon-counting errors. Correlated noise could be introduced by poorly-corrected telluric lines or by unresolved features in the region of the spectra assumed to represent the continuum.

We drew our errors from a multivariate Gaussian with Gaussian weights along the diagonal of the covariance matrix.
At each pixel, we simulated Gaussian random noise using the errors returned by the \spex\ pipeline, which included photon, residual sky, and read noise and which were propagated through the \texttt{Spextool} pipeline. We multiplied the error realization by a Gaussian centered on that pixel with unit area and full width at half maximum (FWHM) equal to the width of the autocorrelation function. To determine the appropriate FWHM, we autocorrelated each order of several spectra of different S/N and found that a Gaussian with a FWHM of 1.5 pixels approximated the width of the autocorrelation function; we used this FWHM for all stars. We did this for each pixel, resulting in an array of overlapping Gaussians of unit area, one centered on each pixel. We then added the contributions from the Gaussians at each pixel, and took the sum at each pixel to be the error on that pixel. 
This effectively spread the error associated with a single pixel over the neighboring pixels according to the autocorrelation function. 

We then re-measured spectral indices (described below), EWs (described in \S\ref{Sec:relation}) and the radial velocity (as described in \S\ref{Sec:rvmethod}). We repeated this process 50 times and calculated the $1\sigma$ confidence intervals, which we took to be the errors on our measurements. 

\begin{figure}
\includegraphics[width=\linewidth]{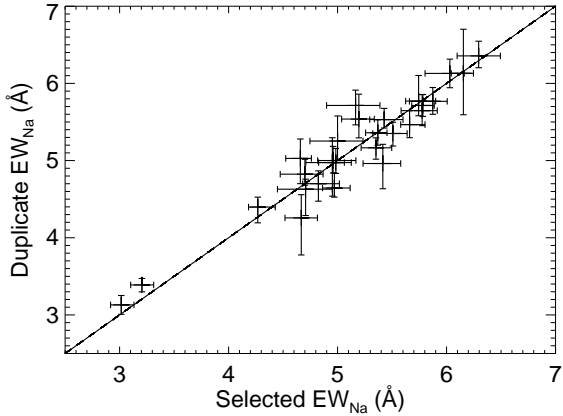}
\caption{We compare $\ewna$ measurements for stars for which we have more than one observation. The horizontal axis shows the $\ewna$ of the selected observation and the vertical axis shows the $\ewna$ of the alternate observation, both in \AA. We also include the $1\sigma$ confidence intervals from 50 trials.
\label{Fig:multobs-line}}
\end{figure}

To assess the accuracy of our error estimates, we considered stars that we observed on two separate occasions, which have different observing conditions and S/N. By comparing independent measurements of the same object, we determined whether our error estimates accurately model the observed differences in the measurements. We used EWs, which we measure by numerically integrating within a defined region, as indicators of M dwarf metallicity (our method is described in detail in \S\ref{Sec:metallicity}). The line of most interest to us is the \ion{Na}{1} line at $2.2\micron$. The median error on $\ewna$ is $0.17\mathrm{\AA}$, typically $5\%$, which was achieved with $\mathrm{S/N}=300$. 92\% of our spectra have S/N in the \K-band greater than 200 and 67\% have errors on $\ewna$ less than $0.2$\AA. In Figure \ref{Fig:multobs-line}, we compare $\ewna$ for stars that were observed multiple times, finding that our method accurately captures the observed errors. 

\begin{figure}
\includegraphics[width=\linewidth]{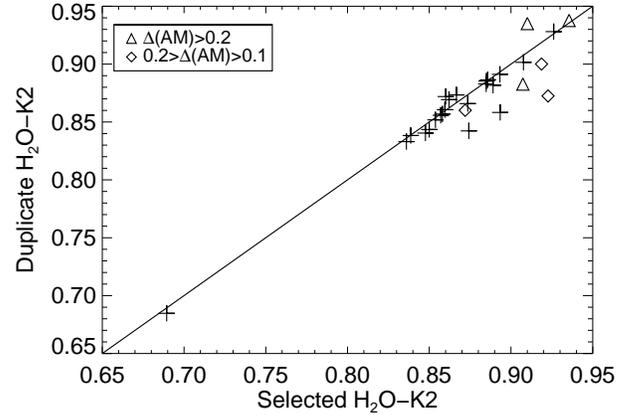}
\caption{We compare measurements of the $\hind$ index for stars which we observed multiple times. On the horizontal axis we show the $\hind$ index of the selected observation and on the vertical axis, the $\hind$ index of the alternate observation. The errors from 50 trials are smaller than the data points. We indicate the cases of significant airmass discrepancies between the science and telluric spectra as triangles (for $\Delta\mathrm{{AM}}>0.2$) and diamonds (for $0.2>\Delta\mathrm{{AM}}>0.1$). The two cases with large discrepancies in the $\hind$ index but for which the science and telluric spectra are closely matching in airmass are instances where the science and telluric observations were separated by more than two hours.
\label{Fig:multobs-hind}}
\end{figure}

We also measure 10 spectral indices (\S\ref{Sec:relation}), including the $\hind$ index, a temperature-sensitive index that measures the curvature of the \K-band by considering the flux level in three \K-band regions (R12). It is defined as: 
\begin{equation}
\hind=\frac{\la2.070-2.090\ra/\la2.235-2.255\ra}{\la2.235-2.255\ra/\la2.360-2.380\ra}
\end{equation}
Angle brackets represent the median flux within the wavelength range indicated, where wavelengths are given in microns. In Figure \ref{Fig:multobs-hind}, we compare measurements of the $\hind$ index for objects which were observed multiple times. Our autocorrelation analysis underestimated the true uncertainties. The largest discrepancies arose when airmass differed by more than 0.2 or time of observation differed by more than two hours (these were not typical occurrences amongst our sample). If using the $\hind$ index for metallicity or temperature measurements, we suggest taking particular care to observe a telluric standard immediately before or after each science observation, and as closely matching in airmass as possible, as described in \citet[]{Vacca2003}.



\section{NIR spectral types}\label{Sec:sptypes}

We determined NIR spectral types by eye for each star using the \K, \H, \J\ and \Z-bands. Our NIR spectral types are based on the spectral typing system defined by \citet{Kirkpatrick1991, Kirkpatrick1995, Kirkpatrick1999}, hereafter the KHM system. We used a custom spectral typing program to match each science spectrum to a library of spectral type standards created from our data (\S\ref{Sec:byeye}-\S\ref{Sec:spstandards}). We considered the differences between our NIR spectral types and other spectral type indicators (\S\ref{Sec:spcompare}) and calibrated a new spectroscopic distance relation using apparent $K_S$ magnitude and either NIR spectral type or the $\hind$ index (\S\ref{Sec:specdist}). 

\subsection{Spectral typing routine}\label{Sec:byeye}

\begin{figure}
\includegraphics[width=0.85\linewidth]{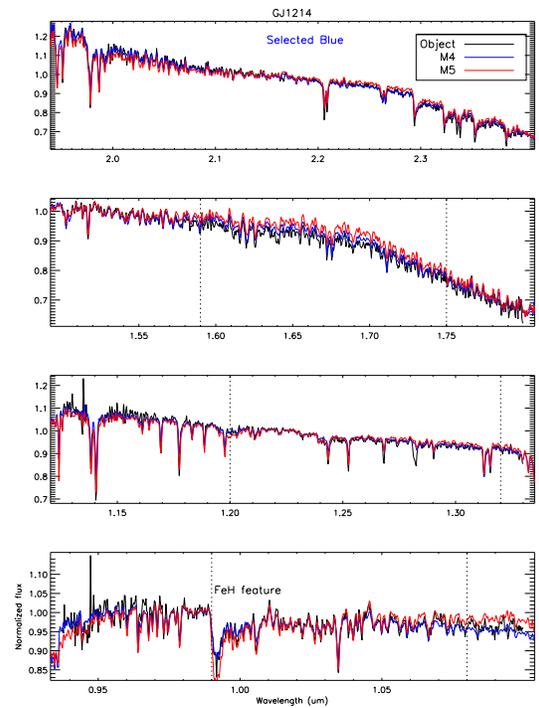}
\caption{An example of the output from our spectral typing routine. We included the \K, \H, \J\ and \Z-bands in our program. We show the object spectrum, in this case GJ 1214, in black. We overplot two spectral standards in blue and red. Dashes indicate FeH bands; only the Wing-Ford band head at $0.99\micron$ is apparent in mid M dwarfs. In this case, we selected the blue spectral standard, M4V, as the best match to the object spectrum. The spectral type from \citet[]{Reid1995} is M4.5V.
\label{Fig:sptyping}}
\end{figure}

\begin{figure}
\includegraphics[width=\linewidth]{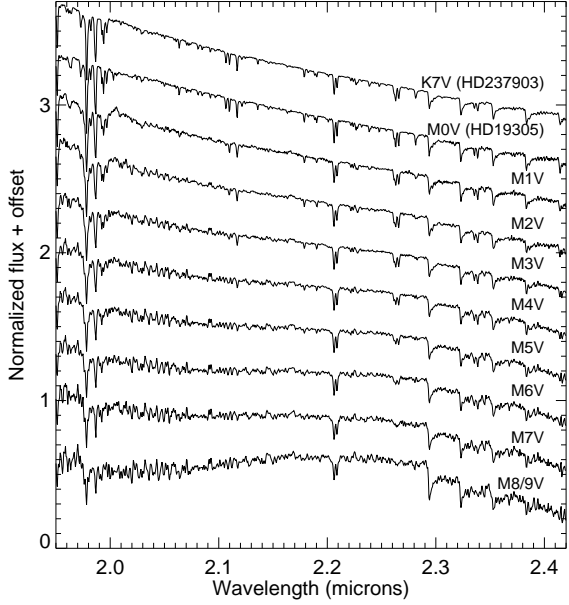}
\caption{Our IRTF spectral sequence from K7V to M9V for the \K\ band. For K7V and M0V, we used the spectral standards from the IRTF library. For the remaining spectral types, we created standards from our observations by median-combining stars of a single spectral type. We were unable to reliably separate M8V and M9V stars and therefore treat them as one spectral category (see \S\ref{Sec:byeye}). In practice, we also could not distinguish between K7V and M0V and assigned these a K7/M0V spectral type. 
\label{Fig:masterK}}
\end{figure}
\begin{figure}
\includegraphics[width=\linewidth]{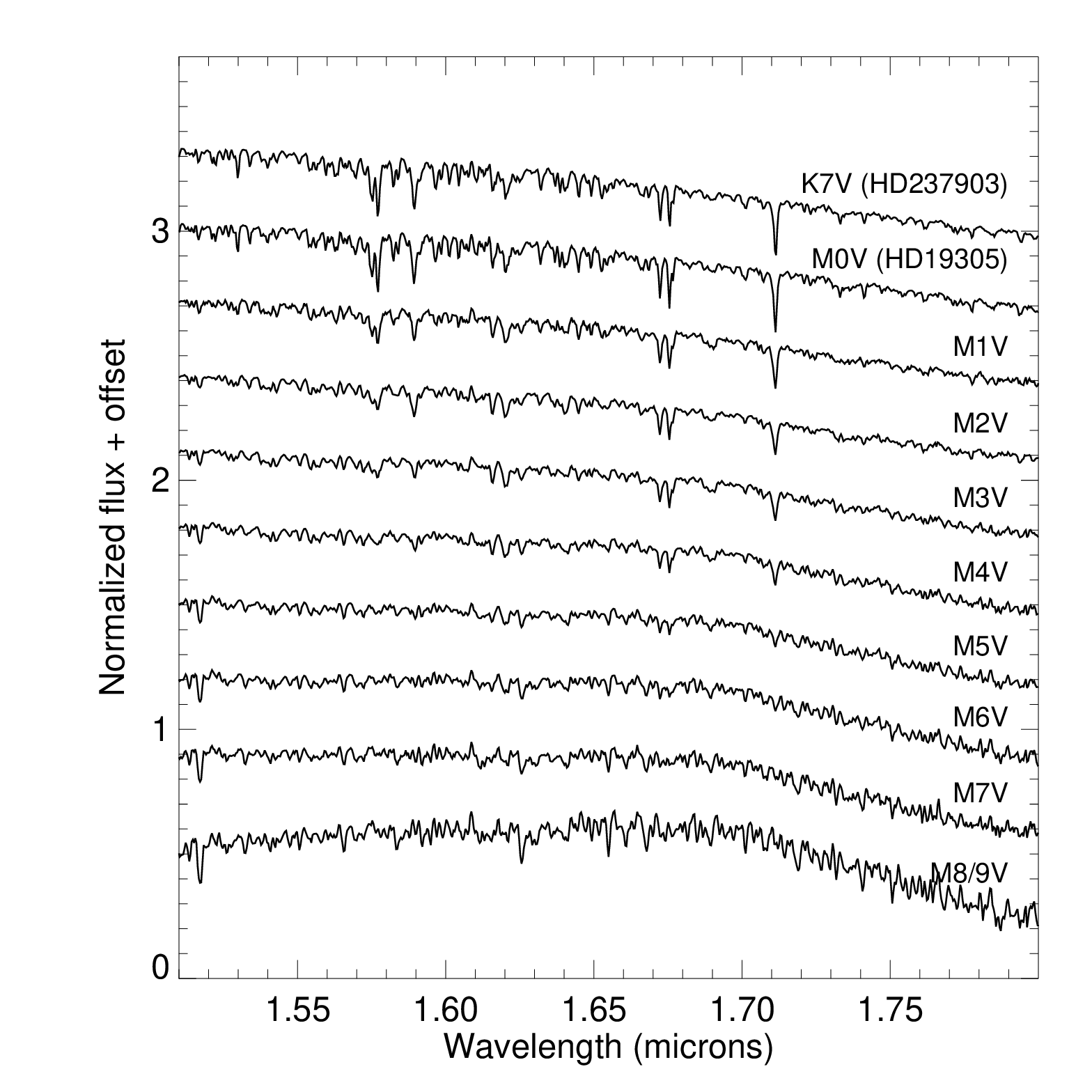}
\caption{Same as in Figure \ref{Fig:masterK} but for the \H\ band.
\label{Fig:masterH}}
\end{figure}
\begin{figure}
\includegraphics[width=\linewidth]{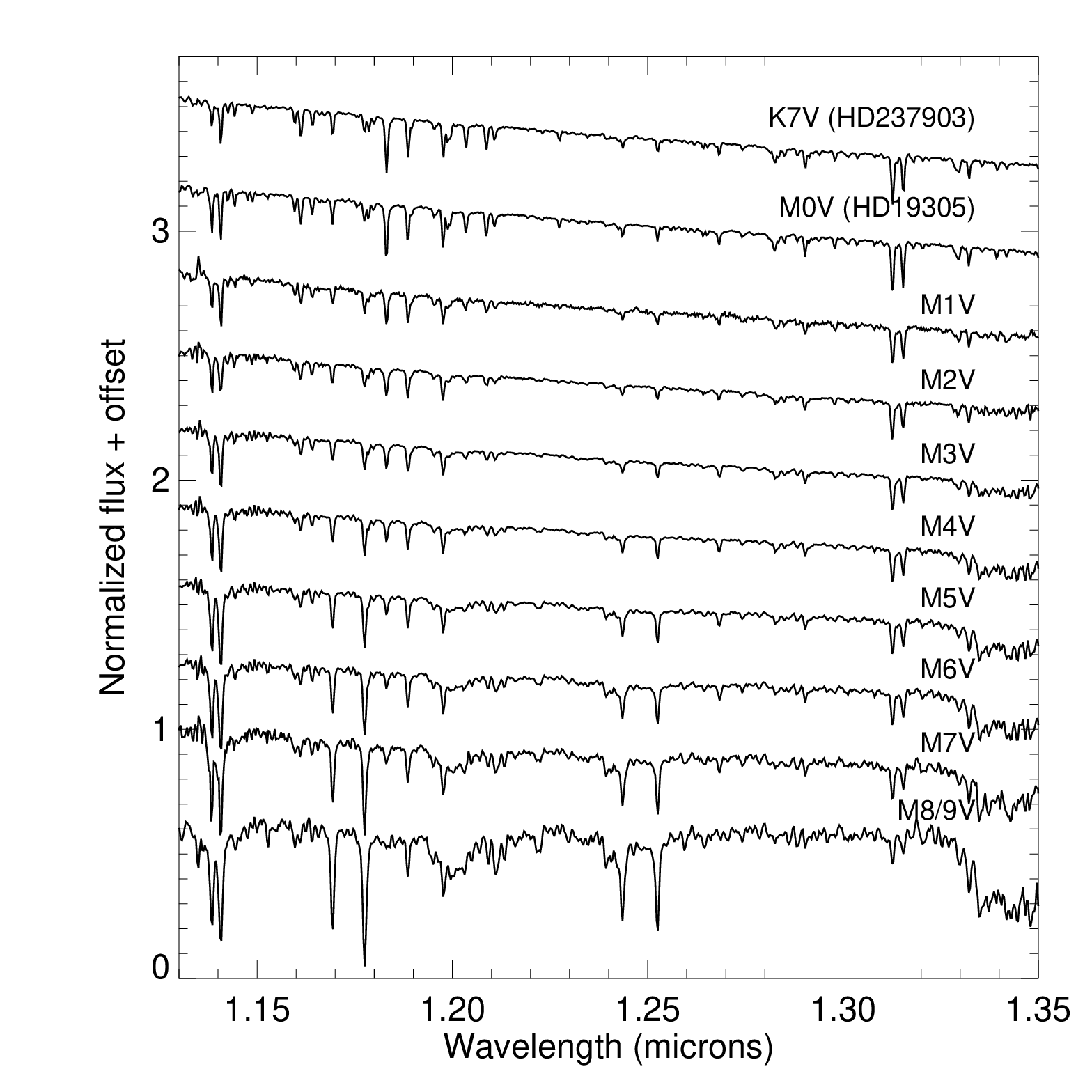}
\caption{Same as in Figure \ref{Fig:masterK} but for the \J\ band.
\label{Fig:masterJ}}
\end{figure}
\begin{figure}
\includegraphics[width=\linewidth]{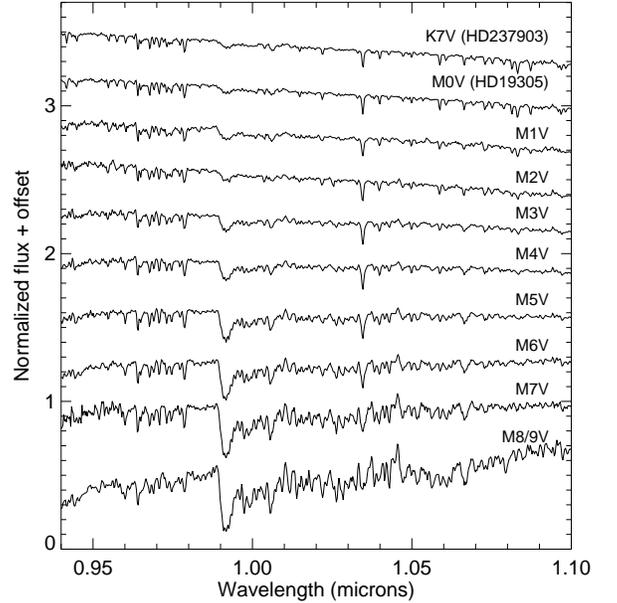}
\caption{Same as in Figure \ref{Fig:masterK} but for the \Z\ band.
\label{Fig:masterZ}}
\end{figure}

We first estimated the spectral type for each star using the relationship between $\hind$ index and spectral type that was presented in R12. We displayed the object spectrum and two spectral standards: the spectral standard with the estimated spectral type and the spectral standard with the spectral type one subtype later. We indicated the FeH bands identified in \citet{Cushing2005} with dashed lines, though the Wing-Ford FeH band at $0.99\micron$ is the only band head apparent across the entire M spectral sequence. FeH is known to be sensitive to spectral type \citep[e.g.][]{Schiavon1997, Cushing2005}. Using a GUI, we checked earlier and later spectral standards as desired, then selected a spectral type for the object. An example is shown in Figure \ref{Fig:sptyping}.

We did not consider half-spectral types. We found the difference between late K dwarfs and M0V stars, and similarly between M8V and M9V stars, to be marginal in the NIR. We used a combined M8V/M9V spectral standard in our program. While K7V and M0V spectral standards were included separately in our spectral typing code, in our later analysis stars we considered a joint K7/M0V spectral class. We took a holistic approach to spectral typing due to the metallicity-dependence of many spectral features. We placed more weight on the redder orders and less weight on features known to be sensitive to metal content (such as the sodium line at $2.2\micron$). Our NIR spectral types are included in Table A1.

\subsection{IRTF spectral standards}\label{Sec:spstandards}

We initially used the M dwarfs in the IRTF spectral library \citep{Rayner2009} as spectral standards, using the KHM spectral standards except for our M0V (HD19305), M3V (AD Leo/Gl 388) and M6V (CN Leo/Gl 406) spectral standards. However, we noted several differences between the strengths of features in the standard spectra and the typical object spectra. In particular, the M4V spectral standard, Gl 213, is metal poor. This is to be expected: \citet[]{Cushing2005} identify Gl 213 as a probable low-metallicity object on the basis of its low Fe, Al, Na, and Ca EWs. By comparison with neighboring spectral standards and using our holistic approach to spectral typing, we were nevertheless able to accurately assess the NIR spectral types of solar metallicity stars. 

To address the concern of spectral standards with extreme metallicities or other unique features, we created our own standard spectra. We assessed the spectral type of all stars observed through the 2012A semester by eye once, using the IRTF spectral library stars as standards. We then median-combine stars of a single spectral type that were within $0.2$ dex of solar metallicity or, for M5V-M7V stars, within $0.1$ dex of solar (see \S\ref{Sec:metallicity} for a description of how we determine metallicities for our stars). There were two stars comprising the M1V spectral standard (with median $\overline{\feh}=0.05$), 10 in M2V ($\overline{\feh}=0.0$), 17 in M3V ($\overline{\feh}=0.02$), 45 in M4V ($\overline{\feh}=0.01$), 48 in M5V ($\overline{\feh}=0.03$), 18 in M6V ($\overline{\feh}=0.04$) and six in M7V ($\overline{\feh}=0.04$). We included all five M8/9V stars observed through the 2012A semester in the M8/M9V spectral standard.  We continued to use the IRTF spectral library standards for K dwarfs and M0V stars. We show our spectral sequence in four IRTF bands, from K7V to M8/9V, in Figs. \ref{Fig:masterK}-\ref{Fig:masterZ}. We then re-classified each star by eye using our new standard spectra.

\subsection{Comparing measures of spectral type}\label{Sec:spcompare}

We first compare our by-eye NIR spectral types to those measured with the $\hind$ index, using the relation in R12. These measures agree to within one spectral type; however, our by-eye spectral types are on average half a spectral type later than those measured using the $\hind$ index. We express M subtype numerically as $\mathrm{Sp}_\mathrm{NIR}$, where positive values are M subtypes (e.g. $\mathrm{Sp}_\mathrm{NIR}=4$ corresponds to M4V) and negative values are K subtypes (e.g. $\mathrm{Sp}_\mathrm{NIR}=-1$ corresponds to K7V and $\mathrm{Sp}_\mathrm{NIR}=-2$ corresponds to K5V). We find: 
\begin{equation}\label{Eq:sptype}
\mathrm{Sp}_\mathrm{NIR} = 25.4 - 24.2\left(\hind\right)
\end{equation}

\begin{figure}
\includegraphics[width=\linewidth]{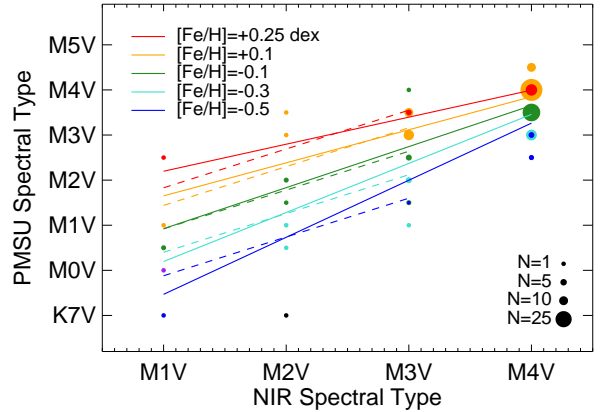}
\caption{Relation between NIR spectral type, metallicity and PMSU spectral type. The horizontal axis is the NIR spectral type determined by eye in this work. The vertical axis is the spectral type from PMSU \citep[]{Reid1995, Hawley1996}, determined from optical spectral features. We represent each bin as a single point, using color to indicate the mean metallicity and size to indicate the number of objects in each bin. In cases where a quarter of the stars fall into a metallicity bin different than the mean, we plot a second data point interior to the first. The area of the interior point relative to the exterior is proportional to the fraction of stars with the second metallicity. Overplotted is our best fitting relation (solid lines). We also include the best fitting linear relation (dashed lines), which extend across the region for which they were calibrated. Contours for our best fits are given by metallicities indicated in the legend and correspond to the colors used for the data points. The metallicity bins used to color data points are:  $-1.0<\feh<-0.6\dex$ (purple),  $-0.6<\feh<-0.4$ (blue),  $-0.4<\feh<-0.2$ (cyan),  $-0.2<\feh<0.0$ (green),  $0.0<\feh<+0.2$ (orange), and $+0.2<\feh<+0.3$ (red). 
\label{Fig:sptypes}}
\end{figure}

Over 100 of our objects have optical spectral types from the Palomar/Michigan State University (PMSU) Survey \citep[][included for comparison in Table A1]{Reid1995, Hawley1996}. The PMSU survey used the depth of the strongest TiO feature in optical M dwarf spectra as the primary indicator of spectral type, and calibrated their relation against nearly 100 spectral classifications on the KHM system.  As in R12, we find a systematic difference between the PMSU spectral types and the NIR spectral types as a function of metallicity, shown in Figure \ref{Fig:sptypes} for stars earlier than M5V.  For M5V stars, there appears to be no clear trend with metallicity.

For early and mid M dwarfs, the NIR spectral type is typically half a spectral type later than the PMSU spectral type, with more metal poor stars being prone to the largest differences between the PMSU and NIR spectral types. We see the same trend with metallicity as in R12: stars that are metal poor were assigned PMSU spectral types that are earlier than the NIR spectral type we assigned.

We calibrated a metallicity-sensitive function relating NIR spectral type to PMSU spectral type, to facilitate joint use of our data. We found that a linear combination of NIR spectral type and metallicity is sufficient only between NIR spectral types M1V and M3V, while a non-linear combination qualitatively explains the trends seen in our data. Our best fitting non-linear relation is given by:
\begin{align}\label{Eq:pmsu}
\mathrm{Sp}_\mathrm{PMSU} = 0.47 + 0.82\left(\mathrm{Sp}_\mathrm{NIR}\right)+4.5\left(\feh\right)\\-0.89\left(\mathrm{Sp}_\mathrm{NIR}\right)\left(\feh\right)
\end{align}
where spectral types are expressed numerically, as described above, and $\feh$ is given in $\dex$. It is valid over NIR spectral types from M1V-M4V and has a scatter of half a subtype.

\subsection{Spectroscopic distances}\label{Sec:specdist}

\begin{figure}
\includegraphics[width=\linewidth]{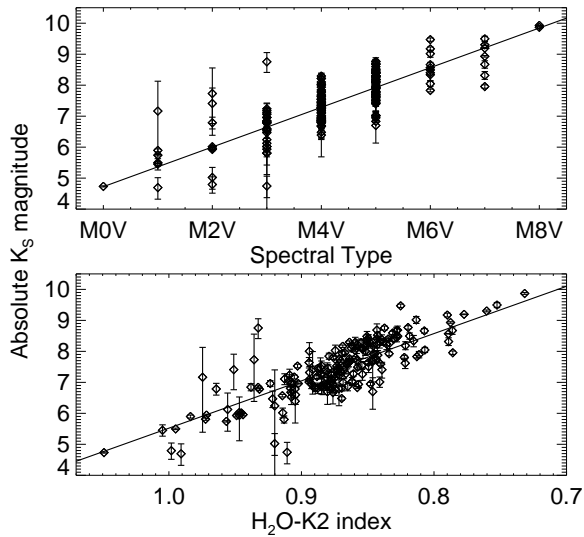}
\caption{Absolute $K_S$ magnitude versus NIR spectral type (top panel) and $\hind$ (bottom panel) for 187 M dwarfs. Overplotted are our best fits. Excluding $5\sigma$ outliers, the standard deviation is $0.30$ magnitudes for the NIR spectral type relation and $0.27$ magnitudes for the $\hind$ relation. The error in the distance inferred by this method is $14\%$.
\label{Fig:specdist}}
\end{figure}

We used NIR spectral type and the $\hind$ index to calibrate a relation with absolute $K_S$ magnitude, using 187 M dwarfs with parallaxes and $K_S$ magnitudes (Figure \ref{Fig:specdist}). We calculated errors on absolute $K_S$ magnitude from the parallax errors, imposing a lower limit of $0.01$ magnitudes (this limit was applied to three stars). We performed a linear least squares fit, using the average of the positive and negative errors on the distance to calculate the $K_S$ magnitude measurement error. The fit is valid for NIR spectral types M0V-M8V or $0.7<\hind<1.06$. Expressing the M subtype numerically, our best fits are:
\begin{align}\label{Eq:specdist}
M_K &= 4.72+0.64\left(\mathrm{Sp}_\mathrm{NIR}\right) \\
&= 20.78-15.26\left(\hind\right)
\end{align}

To estimate the error on the inferred magnitudes and distances, we remove $5\sigma$ outliers and calculate the standard deviation between the measured and inferred absolute magnitudes. Outlier rejection removes four objects for the spectral type relation and three for the $\hind$ relation. The standard deviation is $0.30$ magnitudes for the NIR spectral type relation and $0.27$ magnitudes for the $\hind$ relation, indicating that most of the scatter is intrinsic, rather than due to binning by spectral type. Using standard Gaussian error propagation,
we estimated that the uncertainty in the distances inferred using Equation \ref{Eq:specdist}
is approximately 14\%. Spectroscopic distance estimates based on the $\hind$ index are included for stars in our sample in Table A1. For binaries where only the total magnitude of two components is available, we assume they contribute equally to the luminosity.

\section{Metallicity calibration}\label{Sec:metallicity}

We calibrated our metallicity relation using M dwarfs in CPM pairs with FGK stars, where the primary has a measured metallicity (\S\ref{Sec:primaries}).  Our method of identifying CPM pairs and additional validation using radial velocities and spectroscopic distance estimates is described in \S\ref{Sec:pairs}. We searched the NIR for suitable tracers of metallicity (\S\ref{Sec:relation}) and looked into potential sources of bias (\S\ref{Sec:bias}). We tested our calibration using M dwarf-M dwarf binaries and M dwarfs observed at multiple epochs (\S\ref{Sec:met-check}) and compared measurements from R12 to those from this work (\S\ref{Sec:tspec}).

\subsection{Metallicities of the primary stars}\label{Sec:primaries}

For our potential primary stars, we used FGK stars with metallicities measured by \citet[][hereafter \spocs]{Valenti2005}, \citet[][hereafter \santos]{Santos2004, Santos2005, Santos2011}, \citet[][hereafter \sousa]{Sousa2006, Sousa2008, Sousa2011}, and \citet[]{Bonfils2005}. We use \spocs\ metallicities where available. We also considered those stars with metallicities measured from \citet[]{Sozzetti2009}. \spocs\ and \citet{Sozzetti2009} fit an observed spectrum to a grid of model spectra \citep[]{KuruczR.L.1992}. They reported errors of $0.03$ dex on $\feh$ for measurements of a single spectrum. Work by \santos, \sousa, and \citet{Bonfils2005} used the EWs of iron lines in conjunction with model spectra to measure $\feh$.

\begin{figure}
\includegraphics[width=\linewidth]{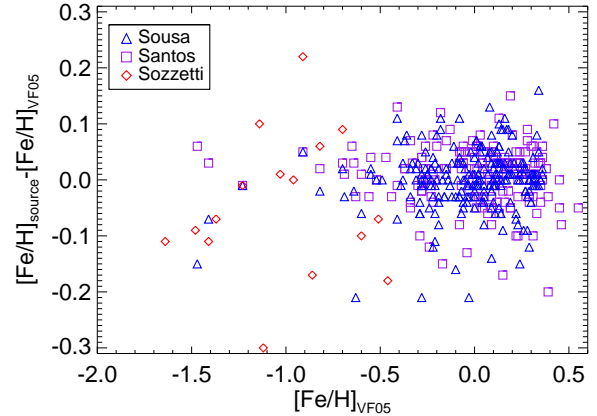}
\caption{Comparison of $\feh$ measurements for single FGK stars from \sousa\ (blue triangles), \santos\ (purple squares) and \citet[][red diamonds]{Sozzetti2009} to \spocs\ $\feh$\ measurements. Metallicities are in dex. We did not use measurements from \citet[]{Sozzetti2009} to calibrate our relation.
\label{Fig:metcheck}}
\end{figure}

We verified that $\feh$\ measurements for FGK stars from different sources are not subject to systematic differences. In Figure \ref{Fig:metcheck}, we compare the $\feh$\ values measured by \sousa, \santos, and \citet[]{Sozzetti2009} to the \spocs\ measurements for single FGK stars, finding the majority of measurements are within $0.1$ dex. The differences between the metallicities from these sources and \spocs\ are $0.00\pm0.05$ for \sousa, $0.00\pm0.06$ for \santos, and $-0.05\pm0.13$ for \citet[]{Sozzetti2009}. Our findings are consistent with those from \citet[]{Sousa2011} and \citet[]{Sozzetti2009}. We did not have a large sample with which to compare $\feh$ measurements from \citet[]{Bonfils2005} and \spocs. However, \citet{Bonfils2005} followed the methods of \citet[]{Santos2004} to measure $\feh$ and found that their work is in agreement.

Out of the 46 M dwarfs in FGK-M CPM pairs with metallicity measurements, there are four M dwarfs for which only a metallicity measurement from \citet{Sozzetti2009} was available: LSPM J0315+0103, LSPM J1208+2147N, LSPM J1311+0936 and PM I16277-0104. These M dwarfs are useful in extending the calibration regime to lower metallicities, but the scatter in their measured metallicities was large enough to be of concern, so we did not use these M dwarfs as part of our final calibration sample. However, we did use these four stars to validate the extrapolation of our calibration to $\feh=-1.0\dex$. 

We used $0.03$ dex divided by the square root of the number of spectra analyzed as the error for \spocs\ metallicity measurements, as described by the authors (typically 1-2 spectra were analyzed in \spocs). Errors for metallicities from \santos, \sousa, and \citet[]{Bonfils2005} were reported individually in the literature. Since the errors were consistent with the scatter we find between \spocs\ and these measurements, we did not further inflate the error bars. 

\subsection{Identification of calibrators}\label{Sec:pairs}

We used calibrators from previous works \citep[]{Bonfils2005,Johnson2009,Schlaufman2010,Terrien2012}, but also identified new calibrators. To locate new FGK-M CPM pairs, we cross-matched the LSPM-North and LSPM-South (L\'epine, private communication) catalogs with themselves and with those stars with measured metallicities from \spocs, \sousa, \santos, or \citet[]{Sozzetti2009}. Our search was subject to the following requirements: the secondary must be within $5\arcmin$, have colors consistent with an M dwarf ($V-J > 2$, $J-K_S > 0.6$ and $H-K_S > 0.1$), and have proper motions within $6\sigma$ of the primary, where the uncertainties were assumed to be those stated in the LSPM catalogs.

We used a $\chi^2$ statistic to identify CPM pairs. The statistic was constructed from the angular separation ($a$), the difference in proper motions ($\Delta\mathrm{PM}=|\boldmath{PM}_1-\boldmath{PM}_2|$), and the difference in distance modulii ($\Delta\mathrm{DM}=\mathrm{DM}_1-\mathrm{DM}_2$). For the distance, we used parallaxes where available, and otherwise used the $M_J$ versus $V-J$ relation \citep{Lepine2005} using the $V-J$ estimates from \citet[]{Lepine2005a}:
\begin{equation}
\chi^2=\left(\frac{a}{2\arcmin}\right)^2+\left(\frac{\Delta\mathrm{PM}}{\sigma_{\Delta\mathrm{PM}}}\right)^2+\left(\frac{\Delta\mathrm{DM}}{\sigma_{\Delta\mathrm{DM}}}\right)^2
\end{equation}
We required $\chi^2<15$ for selection of an object as a candidate binary. 

We note that the $M_J$ values estimated from \citet[]{Lepine2005a} $V-J$ measure were often highly uncertain, because many were derived from photographic estimates of the \V magnitude. Thus, the constraints from requiring a common distance modulus are weak in these cases. Additionally, the LSPM catalogs gave the same proper motion value for many very close systems where separate values could not be measured; our analysis assumed that the proper motions were independently measured.

After gathering our observations, we checked that the RV of the primary was in agreement with our measurement of the RV of the secondary and that the distance to the primary was in agreement with our spectroscopic distance estimate for the secondary. We compared the RV and distance measurements for each calibrator and three stars were immediately obvious as outliers. Two have RVs differing by more than $10\sigma$: Gl 806.1B and CE 226. One has a distance differing by $7.5\sigma$: HD 46375B. (This star is noted on SIMBAD as not being a CPM pair, although in MEarth imaging they do appear to move in tandem). LP 731-76, a mid M dwarf, has the same $K_S$ magnitude as its primary, an early K dwarf, clearly indicating that these are not associated. We did not include these four stars in our final sample of calibrators. While some of these systems may be physically associated, unresolved hierarchical triples, we consider the purity of our sample to be more important than its completeness. 

Two of the remaining calibrators are concerning, but we do not have sufficient cause to exclude them from our sample. LSPM J0045+0015N has a distance estimate of $22 \pc$ (compared to $41 \pc$ for the primary) and an RV of $16 \kms$ (compared to $32 \kms$). 2MASS J03480588+4032226 has a distance estimate of $30 \pc$ (compared to $50 \pc$ for the primary) and an RV of $0 \kms$ (compared to $-10 \kms$); the low proper motion of this object means that the evidence for the physical association of the pair from proper motion alone is weakened. 

We identified 2MASS J17195815-0553043 as a visual double, and a comparison between the National Geographic Society-Palomar Observatory Sky Survey and 2MASS indicates the pair likely has a common proper motion. The distance estimates and radial velocities of the components also support the pair being physical associated. To estimate the distance to 2MASS J17195815-0553043, we assumed the two components had equal magnitudes such that the sum of their fluxes matched the published value. PM I14574-2124W (Gl 570BC) is a known spectroscopic binary, comprised of $0.6\msun$ and $0.4\msun$ components \citep{Forveille1999}. 
As we demonstrate below, the $\na$ line we use to measure metallicity
appears to be only weakly sensitive to temperature over the spectral
type range of our calibration, and therefore the EWs
should not be strongly influenced by the presence of a binary
companion, and this object was not removed from the calibration
sample. To be consistent with our treatment of known and unknown spectroscopic binaries, we use the total magnitude of PM I14574-2124W when estimating its distance.

The M dwarf calibrators and our observations are presented in Tables \ref{Tab:binaries} and \ref{Tab:binaries-measurements}. 46 FGK-M CPM pairs appear in these tables. As previously stated, four of these objects were removed from our final calibration sample because they may not be physically associated. An additional four M dwarfs with measurements of the primary star's metallicity from \citet[]{Sozzetti2009} were not included in the calibration sample, although we used them to validate our calibration to lower metallicities. Two M0V dwarfs were also not included in our final metallicity calibration, as is discussed in subsequent sections. Our final calibration sample therefore consisted of 36 M dwarfs with NIR spectral types from M1V to M5V, with one M7 dwarf,  and metallicities between $-0.7$ and $+0.45 \dex$. The typical calibrator is an M4 or M5 dwarf and has a metallicity within $0.2 \dex$ of solar.

\begin{deluxetable*}{l l l r r l r r l r r r l}  
\setlength{\tabcolsep}{0.001in} 
\tablecaption{\label{Tab:binaries} Observational properties of M dwarf Common Proper Motion pairs}
\tabletypesize{\scriptsize}
\tablecolumns{13}
\tablehead{\colhead{Secondary} & \colhead{RA\tablenotemark{a}} & \colhead{Dec\tablenotemark{a}} & \colhead{$\mathrm{PM}_\mathrm{RA}$} & \colhead{$\mathrm{PM}_\mathrm{Dec}$} & \colhead{Astrom.\tablenotemark{b}} & \colhead{$K_S$\tablenotemark{c}} & \colhead{$d_\mathrm{Sp}$\tablenotemark{d}} & \colhead{Primary} & \colhead{$\mathrm{PM}_\mathrm{RA}$\tablenotemark{e}} & \colhead{$\mathrm{PM}_\mathrm{Dec}$\tablenotemark{e}} & \colhead{$d$\tablenotemark{e}} \\
 & \colhead{(hh:mm:ss)} & \colhead{(dd:mm:ss)} & \colhead{(as/yr)} &  \colhead{(as/yr)} & \colhead{(Ref.)} & \colhead{(mag)} & \colhead{(pc)} &  &  \colhead{(as/yr)} &  \colhead{(as/yr)} & \colhead{(pc)}}  
\startdata
\multicolumn{13}{l}{M dwarfs used to calibrate metallicity relation} \\
\hline
LSPM J0045+0015N      & 00:45:13.58 & $+$00:15:51.0 &  0.207 & $-$0.041 & LS05 &  9.260 &  22 &  HD 4271  &  0.265 & $-$0.051 &   41   \\
Gl  53.1B              & 01:07:38.53 & $+$22:57:20.8 &  0.102 & $-$0.492 & LS05 &  8.673 &  20 &  HD 6660  &  0.099 & $-$0.492 &   20   \\
G 272-119      & 01:54:20.96 & $-$15:43:48.2 &  0.295 & $-$0.137 & S06/SG03 &  9.434 &  38 &  HD 11683  &  0.299 & $-$0.137 &   36    \\
LSPM J0236+0652       &  02:36:15.26 &  $+$06:52:18.0 &  1.813 &  1.447 & LS05 &  6.570 &   6 &  HD 16160  &  1.810 &  1.449 &    7   \\
LSPM J0255+2652W      & 02:55:35.78 & $+$26:52:20.5 &  0.270 & $-$0.191 & LS05 &  8.660 &  20 &  HD 18143  &  0.274 & $-$0.185 &   22 \\
GJ 3195B              & 03:04:43.45 & $+$61:44:09.0 &  0.717 & $-$0.697 & LS05 &  8.103 &  19 &  HD 18757  &  0.721 & $-$0.693 &   22     \\
2MASS J03480588+4032226           & 03:48:05.8  & $+$40:32:22.6   & 0.049 & 0.022 & LG11 &  8.450 &  28 &  HD 23596 &  0.054 &  0.021 &   50   \\
Gl 166C               & 04:15:21.56 & $-$07:39:21.2& $-$2.239 & $-$3.419 & S06/SG03 &  5.962 &   3 &  HD 26965  & $-$2.239 & $-$3.420 &    5     \\
LSPM J0455+0440W          & 04:55:54.46  & $+$04:40:16.4 &  0.136 & $-$0.185 & LS05 &  7.620 &  21 &  HD 31412  &  0.136 & $-$0.185 &   30    \\
LSPM J0528+1231       &  05:28:56.50 &  $+$12:31:53.6 &  0.093 & $-$0.211 & LS05 &  8.790 &  18 &  HD 35956  &  0.087 & $-$0.216 &   28     \\
LSPM J0546+0112           & 05:46:19.38 & $+$01:12:47.2 & $-$0.066 & $-$0.148 & LS05 &  8.800 &  39 &  HD 38529  & $-$0.079 & $-$0.141 &   42   \\
LSPM J0617+0507           &  06:17:10.65 &  $+$05:07:02.3 & $-$0.198 &  0.164 & LS05 &  8.270 &  16 &  HD 43587  & $-$0.195 &  0.165 &   19    \\
PM I06523-0511       & 06:52:18.05 & $-$05:11:24.2 & $-$0.576 & $-$0.011 & LG11 &  5.723 &   7 &  HD 50281  & $-$0.544 & $-$0.003 &    8 \\
Gl 297.2B             & 08:10:34.26 & $-$13:48:51.4 & $-$0.250 &  0.050 & S06/SG03 &  7.418 &  17 &  HD 68146  & $-$0.251 &  0.058 &   22    \\
LSPM J0849+0329W      & 08:49:02.26  & $+$03:29:47.1& $-$0.149 &  0.056 & LS05 &  9.910 &  29 &  HD 75302  & $-$0.148 &  0.060 &   29 \\
LSPM J0852+2818       &  08:52:40.86 &  $+$28:18:59.0 & $-$0.467 & $-$0.238 & LS05 &  7.670 &  11 &  HD 75732  & $-$0.485 & $-$0.234 &   12   \\
Gl 376B             & 10:00:50.23 & $+$31:55:45.2 & $-$0.529\tablenotemark{f}  & $-$0.429\tablenotemark{f}  & 2MASS &  9.275 &  11 &  HD 86728  & $-$0.529 & $-$0.429 &   14   \\
LSPM J1248+1204       & 12:48:53.45 & $+$12:04:32.7 &  0.225 & $-$0.128 & LS05 & 10.570 &  36 &  HD 111398  &  0.234 & $-$0.141 &   36  \\
Gl 505B               & 13:16:51.54 & $+$17:00:59.9 &  0.632 & $-$0.261 & LS05 &  5.749 &  10 &  HD 115404  &  0.631 & $-$0.261 &   11     \\
Gl 544B               & 14:19:35.83 & $-$05:09:08.1 & $-$0.633 & $-$0.122 & S06/SG03 &  9.592 &  23 &  HD 125455  & $-$0.632 & $-$0.122 &   20    \\
PM I14574-2124W     & 14:57:26.51 & $-$21:24:40.6 &  0.987 & $-$1.667 & LG11 &  3.802 &   3: &  HD 131977  &  1.034 & $-$1.726 &    5 \\ 
LSPM J1535+6005E          &  15:35:25.69 &  $+$60:05:00.6 &  0.166 & $-$0.160 & LS05 &  8.410 &  15 &  HD 139477  &  0.171 & $-$0.163 &   19  \\
LSPM J1604+3909W      & 16:04:50.85 & $+$39:09:36.1 & $-$0.547 &  0.055 & LS05 &  9.160 &  18 &  HD 144579  & $-$0.572 &  0.052 &   14     \\
PM I17052-0505       & 17:05:13.81 & $-$05:05:38.7 & $-$0.921 & $-$1.128 & LG11 &  5.975 &   8 &  HD 154363  & $-$0.917 & $-$1.138 &   10     \\
2MASS J17195815-0553043A\tablenotemark{g} & 17:19:58.15J & $-$05:53:04.5J &  0.049J & $-$0.182J & LS05 & 10.385J &  55: & HD 156826 &  0.045 &  $-$0.194 &   53     \\
2MASS J17195815-0553043B\tablenotemark{g} & 17:19:58.15J & $-$05:53:04.5J &  0.049J & $-$0.182J & LS05 & 10.385J &  41: &  HD 156826 &  0.045 &  $-$0.194 &   53     \\
LSPM J1800+2933NS     &  18:00:45.43 &  $+$29:33:56.8 & $-$0.128 &  0.169 & LS05 &  8.230 &  24 & HD 164595 &  $-$0.139 &  0.173 &   28     \\
PM I19321-1119       & 19:32:08.11 & $-$11:19:57.3 &  0.237 &  0.026 & LG11 &  8.706 &  18 &  HD 183870  &  0.235 &  0.018 &   18     \\
Gl 768.1B             & 19:51:00.67 & $+$10:24:40.1 &  0.240 & $-$0.135 & 2MASS &  8.012 &  15 &  HD 187691  &  0.240 & $-$0.135 &   19     \\
LSPM J2003+2951       &  20:03:26.58 &  $+$29:51:59.4 &  0.689 & $-$0.515 & LS05 &  8.710 &  14 &  HD 190360  &  0.684 & $-$0.524 &   17     \\
LSPM J2011+1611E      & 20:11:13.26 & $+$16:11:08.0 & $-$0.432 &  0.399 & LS05 &  8.880 &  16 &  HD 191785  & $-$0.413 &  0.398 &   20     \\
LSPM J2040+1954       &  20:40:44.52 &  $+$19:54:03.2 &  0.107 &  0.312 & LS05 &  7.420 &  12 &  HD 197076A  &  0.118 &  0.310 &   19     \\
LSPM J2231+4509       & 22:31:06.51 & $+$45:09:44.0 & $-$0.167 &  0.027 & LS05 &  9.500 &  37 &  HD 213519  & $-$0.174 &  0.038 &   43     \\
Gl 872B               & 22:46:42.34 & $+$12:10:20.9 &  0.234 & $-$0.492 & LS05 &  7.300 &  14 &  HD 215648  &  0.233 & $-$0.492 &   16     \\
LSPM J2335+3100E          & 23:35:29.47  & $+$31:00:58.5 &  0.548 &  0.256 & LS05 &  8.850 &  24 & HD 221830 &  0.539 &  0.254 &   32    \\
HD 222582B       & 23:41:45.14 & $-$05:58:14.8 & $-$0.148 & $-$0.117 & S06/SG03 &  9.583 &  30 &  HD 222582  & $-$0.145 & $-$0.111 &   41    \\
\hline
\multicolumn{9}{l}{M0 dwarfs in a CPM pair not used in our metallicity calibration} \\
\hline
Gl 282B       & 07:40:02.90 & $-$03:36:13.3 &  0.067 & $-$0.286 & H00 &  5.568 &  13 &  HD 61606  &  0.070 & $-$0.278 &   14   \\
LSPM J1030+5559           &  10:30:25.31 &  $+$55:59:56.8 & $-$0.181 & $-$0.034 & LS05 &  5.360 &  13 &  HD 90839  & $-$0.178 & $-$0.033 &   12   \\
\hline
\multicolumn{13}{l}{M dwarfs in a CPM pair where the primary has a metallicity measurement from \citet{Sozzetti2009}} \\
\hline
LSPM J0315+0103       & 03:15:00.922 & $+$01:03:08.2 &  0.362 &  0.118 & LS05 & 10.85 &  77 & G 77-35 & 0.362 & 0.118 & 79  \\
LSPM J1208+2147N          & 12:08:55.378 & $+$21:47:31.6 & $-$0.439 &  0.037 &  LS05 & 10.38 &  83 &  G 59-1 & $-$0.397 & $0.036$  &  113:  \\
LSPM J1311+0936           & 13:11:22.445 & $+$09:36:13.1 & $-$0.517 &  0.269 & LS05 &  8.86 &  55 & G 63-5 & $-$0.521 & $0.269$ & 61    \\
PM  I16277-0104       & 16:27:46.699 & $-$01:04:15.4 & $-$0.340 & $-$0.106 & LS05 & 10.57 &  54 & G 17-16 &  $-$0.347 & $-0.102$   & 62:    \\
\hline
\multicolumn{13}{l}{M dwarfs in a CPM pair that may not to be physically associated} \\
\hline
HD 46375B             & 06:33:12.10 & $+$05:27:53.1 &  0.114 & $-$0.097 & 2MASS &  7.843 &  11 &  HD 46375  &  0.114 & $-$0.097 &   33       \\
CE 226      & 10:46:33.27 & $-$24:35:11.2 & $-$0.141\tablenotemark{f}  & $-$0.109\tablenotemark{f}  & 2MASS &  9.447 &  31 &  HD 93380  & $-$0.141 & $-$0.109 &   20 \\
LP 731-76     & 10:58:27.99 & $-$10:46:30.5 & $-$0.201 & $-$0.094 & S06/SG03 &  8.640 &  14 &  BD-103166  & $-$0.186 & $-$0.005 & 25:\tablenotemark{h}     \\
Gl 806.1B             & 20:46:06.42 & $+$33:58:06.2 &  0.356 &  0.330 & MEarth &  8.7:\tablenotemark{i} & 19: &  HD 197989  &  0.356 &  0.330 &   22     \\
\enddata
\tablerefs{\citet[][H00]{HogE.2000}; \citet[][SG03]{Salim2003}; \citet[LS05]{Lepine2005a}; \citet[S06]{Skrutskie2006};  \citet[][LG11]{Lepine2011}}
\tablenotetext{a}{Positions are given in the International Celestial Reference System (ICRS), and have been corrected to epoch 2000.0 where necessary assuming the proper motions given in the table.}
\tablenotetext{b}{Astrometry references. If one reference is provided, it applies to both position and proper motion; if two are provided, the first is for position and the second for proper motion.}
\tablenotetext{c}{Apparent $K_S$ magnitudes are from S06.}
\tablenotetext{d}{Errors on the distance estimates are 14\%.}
\tablenotetext{e}{Proper motions and distances for primary stars are from Hipparcos \citep{VanLeeuwen2007} except when otherwise noted.}
\tablenotetext{f}{For CE 226 and Gl 376B, the Hipparcos proper motion for the primary was found to be a better match to the observed motion of the secondary from 2MASS to recent epoch MEarth imaging than the proper motion given in \citet[][for CE 226]{Ruiz2001} or in LSPM-North (for Gl 376B). In these cases, the Hipparcos value has been adopted in the table.}  
\tablenotetext{g}{{L}\'{e}pine, private communication. We resolved this object as a binary. An appended ``J'' indicates a measurement that was derived for the components jointly. We assume the two components contribute equally to the luminosity in order to estimate their spectroscopic distances.}
\tablenotetext{h}{No parallax was available for the primary. Its distance was estimated assuming an absolute $K_S$ magnitude of 6, typical for an early K dwarf.}
\tablenotetext{i}{No $K_S$ magnitude could be found for Gl 806.1B. We estimated a rough magnitude from 2MASS Atlas images using a 4 pixel aperture radius (this value was chosen to reduce contamination from nearby stars), and applied an aperture correction of 0.04 magnitudes, derived from stars of similar K magnitude elsewhere in the field.}
\end{deluxetable*}


 \begin{deluxetable*}{l l r r r r r l r r l} 
\setlength{\tabcolsep}{0.0001in} 
\tablecaption{\label{Tab:binaries-measurements} Measured properties of M dwarf CPM pairs}
\tabletypesize{\scriptsize}
\tablecolumns{9}
\tablehead{\colhead{Secondary} & \colhead{SpT} & \colhead{$\ewna$} & \colhead{$\mathrm{EW}_\mathrm{Ca}$}& \colhead{$\hind$}& \colhead{$\feh$} & \multicolumn{2}{c}{ $\feh_\mathrm{prim}$\tablenotemark{a}}  & \colhead{$\mathrm{RV}_\mathrm{sec}$}  & \multicolumn{2}{c}{$\mathrm{RV}_\mathrm{prim}$}  \\
& \colhead{NIR} & \colhead{(\AA)} & \colhead{(\AA)} & & \colhead{(dex)} & \colhead{(dex)} & \colhead{Ref.} & \colhead{(km/s)}  & \colhead{(km/s)} &\colhead{Ref.}}
\startdata
\multicolumn{9}{l}{M dwarfs used to calibrate metallicity relation} \\
\hline
LSPM  J0045-0015N & M4 & $5.24 \pm 0.15$ & $3.41 \pm 0.14$ & $0.868 \pm 0.005$ & $+0.08 \pm 0.12$ & $+0.02 \pm 0.03$ & VF05 & $   16 \pm    5$ &   $32.5$ & VF05 \\ 
Gl 53.1B & M4 & $6.19 \pm 0.16$ & $3.63 \pm 0.14$ & $0.894 \pm 0.005$ & $+0.22 \pm 0.12$ & $+0.07 \pm 0.12$ & B05 & $   16 \pm    5$ &    $7.0$ & Chub10 \\ 
G272-119 & M3 & $4.13 \pm 0.15$ & $3.20 \pm 0.15$ & $0.937 \pm 0.005$ & $-0.17 \pm 0.13$ & $-0.21 \pm 0.03$ & Sou06 & $   11 \pm    5$ &   $-1.2$ & VF05 \\ 
LSPM  J0236-0652 & M4 & $3.96 \pm 0.14$ & $2.49 \pm 0.17$ & $0.866 \pm 0.005$ & $-0.22 \pm 0.13$ & $-0.12 \pm 0.02$ & VF05 & $   30 \pm    6$ &   $26.8$ & VF05 \\ 
LSPM  J0255-2652W & M4 & $6.27 \pm 0.17$ & $3.83 \pm 0.18$ & $0.897 \pm 0.005$ & $+0.23 \pm 0.12$ & $+0.28 \pm 0.03$ & VF05 & $   33 \pm    5$ &  $ 32.5$ & VF05 \\ 
GJ 3195B & M3 & $3.90 \pm 0.18$ & $3.29 \pm 0.17$ & $0.924 \pm 0.005$ & $-0.24 \pm 0.13$ & $-0.31 \pm 0.04$ & B05 & $   -1 \pm    5$ &   $-6.8$ & VF05 \\ 
2MASS J03480588+4032226 & M2 & $8.07 \pm 0.15$ & $5.76 \pm 0.15$ & $0.958 \pm 0.005$ & $+0.29 \pm 0.12$ & $+0.22 \pm 0.03$ & VF05 & $    0 \pm    5$ &  $-10.6$ & VF05 \\ 
Gl 166C & M5 & $3.99 \pm 0.16$ & $2.13 \pm 0.21$ & $0.835 \pm 0.005$ & $-0.21 \pm 0.13$ & $-0.28 \pm 0.02$ & VF05 & $  -37 \pm    6$ &  $-42.3$ & VF05 \\ 
\ldots &  \ldots & \ldots &  \ldots & \ldots & \ldots & $-0.33 \pm 0.06 $&  B05   & \ldots & \ldots & \ldots \\ 
LSPM  J0455-0440W & M3 & $5.60 \pm 0.15$ & $4.84 \pm 0.20$ & $0.965 \pm 0.005$ & $+0.15 \pm 0.12$ & $+0.05 \pm 0.03$ & VF05 & $   46 \pm    5$ &  $ 47.7 $& VF05 \\ 
LSPM  J0528-1231 & M4 & $5.16 \pm 0.20$ & $2.97 \pm 0.21$ & $0.870 \pm 0.005$ & $+0.07 \pm 0.13$ & $-0.22 \pm 0.03$ & VF05 & $   17 \pm    5$ &  $ 17.3$ & VF05 \\ 
LSPM  J0546-0112 & M1 & $7.24 \pm 0.17$ & $5.19 \pm 0.20$ & $0.982 \pm 0.005$ & $+0.30 \pm 0.12$ & $+0.45 \pm 0.03$ & VF05 & $   28 \pm    5$ &  $ 30.2$ & VF05 \\ 
\ldots &  \ldots & \ldots &  \ldots & \ldots & \ldots &  $+0.40 \pm 0.06$  &  San04  & \ldots & \ldots & \ldots  \\ 
LSPM  J0617-0507 & M4 & $5.23 \pm 0.11$ & $3.31 \pm 0.17$ & $0.891 \pm 0.005$ & $+0.08 \pm 0.12$ & $-0.04 \pm 0.03$ & VF05 & $   11 \pm    5$ &   $12.7$ & VF05 \\ 
PM I06523-0511 & M2 & $4.61 \pm 0.07$ & $4.17 \pm 0.10$ & $0.953 \pm 0.005$ & $-0.05 \pm 0.12$ & $+0.14 \pm 0.03$ & VF05 & $   -5 \pm    5$ &   $-5.4$ & VF05 \\ 
Gl 297.2B & M2 & $4.89 \pm 0.23$ & $4.15 \pm 0.26$ & $0.953 \pm 0.005$ & $+0.01 \pm 0.13$ & $-0.09 \pm 0.09$ & B05 & $   30 \pm    5$ &   $37.7$ & VF05 \\ 
LSPM  J0849-0329W & M4 & $5.05 \pm 0.21$ & $3.23 \pm 0.22$ & $0.861 \pm 0.005$ & $+0.05 \pm 0.13$ & $+0.10 \pm 0.03$ & VF05 & $   12 \pm    5$ &   $10.8$ & VF05 \\ 
LSPM  J0852-2818 & M4 & $7.53 \pm 0.19$ & $3.60 \pm 0.24$ & $0.882 \pm 0.005$ & $+0.30 \pm 0.12$ & $+0.31 \pm 0.01$ & VF05 & $   31 \pm    5$ &  $ 27.8$ & VF05 \\ 
\ldots &  \ldots & \ldots &  \ldots & \ldots & \ldots &  $+0.33 \pm 0.07$ &  San04    & \ldots & \ldots & \ldots  \\ 
Gl 376B & M7 & $6.56 \pm 0.26$ & $1.74 \pm 0.24$ & $0.776 \pm 0.005$ & $+0.26 \pm 0.12$ & $+0.20 \pm 0.02$ & VF05 & $   52 \pm    5$ &  $ 56.0$ & Mas08 \\ 
LSPM  J1248-1204 & M5 & $4.46 \pm 0.22$ & $2.70 \pm 0.21$ & $0.854 \pm 0.005$ & $-0.09 \pm 0.13$ & $+0.08 \pm 0.03$ & VF05 & $    8 \pm    5$ &    $3.5$ & VF05 \\ 
Gl 505B & M1 & $3.77 \pm 0.08$ & $3.84 \pm 0.11$ & $0.995 \pm 0.005$ & $-0.27 \pm 0.12$ & $-0.25 \pm 0.05$ & B05 & $    1 \pm    5$ &   $ 8.5$ & C12 \\ 
Gl 544B & M5 & $4.78 \pm 0.27$ & $2.45 \pm 0.31$ & $0.855 \pm 0.005$ & $-0.01 \pm 0.13$ & $-0.18 \pm 0.03$ & VF05 & $    6 \pm    7$ &   $-9.5$ & VF05 \\ 
\ldots &  \ldots & \ldots &  \ldots & \ldots & \ldots & $-0.20 \pm 0.19$ &  B05    & \ldots & \ldots & \ldots \\ 
PM I14574-2124W  & M2 & $5.31 \pm 0.23$ & $4.56 \pm 0.22$ & $0.981 \pm 0.005$ & $+0.10 \pm 0.13$ & $+0.12 \pm 0.02$ & VF05 & $   25 \pm    5$ &   $26.0$ & VF05 \\ 
\ldots &  \ldots & \ldots &  \ldots & \ldots & \ldots  & $+0.07 \pm 0.10$ &  San05    & \ldots & \ldots & \ldots  \\ 
LSPM  J1535-6005E & M5 & $5.38 \pm 0.08$ & $3.94 \pm 0.10$ & $0.877 \pm 0.005$ & $+0.11 \pm 0.12$ & $+0.11 \pm 0.03$ & VF05 & $   -4 \pm    5$ &   $-8.3$ & VF05 \\ 
LSPM  J1604-3909W & M5 & $3.03 \pm 0.20$ & $1.31 \pm 0.16$ & $0.849 \pm 0.005$ & $-0.52 \pm 0.15$ & $-0.69 \pm 0.03$ & VF05 & $  -64 \pm    5$ &  $-59.0$ & VF05 \\ 
PM I17052-0505 & M3 & $3.27 \pm 0.13$ & $3.09 \pm 0.15$ & $0.940 \pm 0.005$ & $-0.44 \pm 0.14$ & $-0.62 \pm 0.04$ & Sou06 & $   24 \pm    6$ &   $33.6$ & VF05 \\ 
2MASS J17195815-0553043A & M4 & $4.17 \pm 0.52$ & $1.86 \pm 0.65$ & $0.842 \pm 0.005$ & $-0.16 \pm 0.18$ & $-0.13 \pm 0.03$ & VF05 & $  -23 \pm    5$ &  $-32.3$ & VF05 \\ 
2MASS J17195815-0553043B & M5 & $4.02 \pm 0.33$ & $2.57 \pm 0.27$ & $0.877 \pm 0.005$ & $-0.20 \pm 0.15$ & $-0.13 \pm 0.03$ & VF05 & $  -25 \pm    6$ &  $-32.3$ & VF05 \\ 
LSPM  J1800-2933NS & M2 & $4.78 \pm 0.19$ & $3.86 \pm 0.18$ & $0.949 \pm 0.005$ & $-0.01 \pm 0.13$ & $-0.06 \pm 0.03$ & VF05 & $    7 \pm    5$ &    $2.4$ & VF05 \\ 
PM I19321-1119 & M5 & $4.70 \pm 0.26$ & $3.50 \pm 0.25$ & $0.880 \pm 0.005$ & $-0.03 \pm 0.13$ & $+0.05 \pm 0.03$ & VF05 & $  -47 \pm    5$ &  $-48.3$ & VF05 \\ 
\ldots &  \ldots & \ldots &  \ldots & \ldots & \ldots &  $-0.07 \pm 0.03$ &  Sou06    & \ldots & \ldots & \ldots  \\ 
Gl 768.1B & M4 & $5.07 \pm 0.30$ & $3.35 \pm 0.27$ & $0.896 \pm 0.005$ & $+0.05 \pm 0.13$ & $+0.16 \pm 0.02$ & VF05 & $    3 \pm    5$ &    $1.4$ & VF05 \\ 
\ldots &  \ldots & \ldots &&& \ldots &   $+0.07 \pm 0.12$ &  B05    & \ldots & \ldots & \ldots \\ 
LSPM  J2003-2951 & M5 & $5.36 \pm 0.21$ & $2.81 \pm 0.15$ & $0.847 \pm 0.005$ & $+0.10 \pm 0.13$ & $+0.21 \pm 0.03$ & VF05 & $  -40 \pm    5$ &  $-44.8$ & VF05 \\ 
LSPM  J2011-1611E & M5 & $3.71 \pm 0.18$ & $1.97 \pm 0.18$ & $0.852 \pm 0.005$ & $-0.29 \pm 0.13$ & $-0.15 \pm 0.03$ & VF05 & $  -45 \pm    5$ &  $-49.0$ & VF05 \\ 
LSPM  J2040-1954 & M3 & $3.97 \pm 0.13$ & $3.00 \pm 0.15$ & $0.913 \pm 0.005$ & $-0.21 \pm 0.12$ & $-0.09 \pm 0.03$ & VF05 & $  -33 \pm    5$ &  $-35.2$ & VF05 \\ 
LSPM  J2231-4509 & M3 & $4.89 \pm 0.22$ & $3.34 \pm 0.29$ & $0.928 \pm 0.005$ & $+0.01 \pm 0.13$ & $-0.00 \pm 0.03$ & VF05 & $  -29 \pm    5$ &  $-31.5$ & VF05 \\ 
Gl 872B & M3 & $4.01 \pm 0.25$ & $3.16 \pm 0.26$ & $0.939 \pm 0.005$ & $-0.20 \pm 0.14$ & $-0.22 \pm 0.01$ & VF05 & $    0 \pm    5$ &-4.5& VF05 \\ 
\ldots &  \ldots & \ldots &  \ldots & \ldots & \ldots  & $-0.36 \pm 0.11$ &  B05    & \ldots & \ldots & \ldots   \\ 
LSPM  J2335-3100E & M4 & $3.09 \pm 0.15$ & $2.42 \pm 0.19$ & $0.904 \pm 0.005$ & $-0.50 \pm 0.14$ & $-0.40 \pm 0.03$ & VF05 & $ -110 \pm    8$ & $-111.8$ & VF05 \\ 
HD 222582B & M3 & $5.03 \pm 0.17$ & $2.97 \pm 0.15$ & $0.892 \pm 0.005$ & $+0.04 \pm 0.12$ & $-0.03 \pm 0.02$ & VF05 & $   21 \pm    5$ &   $12.6$ & VF05 \\ 
\ldots &  \ldots & \ldots &  \ldots & \ldots & \ldots &  $+0.05 \pm 0.05$  &  San04   & \ldots & \ldots & \ldots \\ 
\ldots &  \ldots & \ldots &  \ldots & \ldots & \ldots &  $-0.01 \pm 0.01$ &  Sou06   & \ldots & \ldots & \ldots \\ 
\hline
\multicolumn{9}{l}{M0 dwarfs in a CPM pair not used in our metallicity calibration} \\
\hline
Gl 282B & M0 & $3.85 \pm 0.12$ & $4.36 \pm 0.12$ & $1.044 \pm 0.005$ & $-0.25 \pm 0.13$ & $+0.07 \pm 0.03$ & VF05 & $  -20 \pm    5$ &  $-17.6$ & VF05 \\ 
\ldots &  \ldots & \ldots &  \ldots & \ldots & \ldots &  $+0.01 \pm 0.08$  &  San05   & \ldots & \ldots & \ldots  \\ 
LSPM  J1030-5559 & M0 & $3.56 \pm 0.18$ & $4.13 \pm 0.19$ & $1.049 \pm 0.005$ & $-0.34 \pm 0.14$ & $-0.07 \pm 0.02$ & VF05 & $   10 \pm    5$ &    $9.4$ & VF05 \\ 
\hline
\multicolumn{9}{l}{M dwarfs in a CPM pair where the primary has a metallicity measurement from \citet{Sozzetti2009}} \\
\hline
LSPM  J0315-0103       & M2  & $2.09 \pm 0.21$ & $1.98 \pm 0.27$ & $0.942 \pm 0.005$ & $-0.89 \pm 0.20$ & $-0.77$ & Soz09 & $   87 \pm    5$ & $88.1$ & L02  \\ 
LSPM J1208-2147N          & M2  & $2.54 \pm 0.17$ & $1.97 \pm 0.25$ & $0.984 \pm 0.005$ & $-0.70 \pm 0.17$ & $-1.05$ & Soz09 & $   -3 \pm    7$ & $-9.9$ & L02  \\ 
LSPM J1311-0936           & M0  & $2.90 \pm 0.16$ & $3.10 \pm 0.16$ & $1.025 \pm 0.005$ & $-0.56 \pm 0.15$ & $-0.62$ & Soz09 & $   27 \pm    5$ & $26.8$ & L02  \\ 
PM I16277-0104       & M3  & $2.98 \pm 0.22$ & $2.01 \pm 0.45$ & $0.911 \pm 0.005$ & $-0.54 \pm 0.22$ & $-0.87$ & Soz09 & $ -158 \pm    5$ &  $-162.4$ & L02  \\ 
\hline
\multicolumn{9}{l}{M dwarfs in a CPM pair that may not to be physically associated} \\
\hline
HD 46375B & M1 & $6.62 \pm 0.19$ & $4.97 \pm  0.21$ & $0.988 \pm 0.005$ & $+0.26 \pm  0.12$ & $+0.25 \pm 0.03$ & VF05 & $    0 \pm    5$ &   $-0.4$ & VF05 \\ 
CE 226 & M4 & $3.79 \pm 0.17$ & $2.17 \pm  0.24$ & $0.905 \pm 0.005$ & $-0.27 \pm  0.13$ & $-0.72 \pm 0.03$ & Sou06 & $  -15 \pm    5$ &   $46.5$ & VF05 \\ 
LP 731-76 & M5 & $6.04 \pm 0.17$ & $3.09 \pm  0.15$ & $0.853 \pm 0.005$ & $+0.21 \pm  0.12$ & $+0.38 \pm 0.03$ & VF05 & $   11 \pm    5$ &   $27.2$ & VF05 \\ 
\ldots &  \ldots & \ldots &  \ldots & \ldots & \ldots & $+0.35 \pm 0.05$ &  San05    & \ldots & \ldots & \ldots \\ 
Gl 806.1B & M4 & $3.93 \pm 0.41$ & $3.20 \pm  0.48$ & $0.895 \pm 0.005$ & $-0.23 \pm  0.17$ & $-0.05 \pm 0.13$ & B05 & $   -8 \pm    5$ &   $44.9$ & VF05 \\ 
\enddata
\tablerefs{\citet[][VF05]{Valenti2005};  \citet[][B05]{Bonfils2005}; \citet[][Mal10]{Maldonado2010}; \citet[][Sou06]{Sousa2006}; \citet[][San04]{Santos2004}; \citet[][San05]{Santos2005}; \citet[][Mas08]{Massarotti2008}; \citet[][C12]{Chubak2012}}
\tablenotetext{a}{Reference for published metallicity of the primary star. If more than one value is available, alternative values are provided in the following row(s). Values from the SPOCS catalog (VF05) are preferred.}
\end{deluxetable*}

\subsection{Empirical metallicity calibration}\label{Sec:relation}

We looked for combinations of spectral features that are good tracers of $\feh$. Based on the lines listed in \citet[]{Cushing2005} and \citet[]{Covey2010}, we identified \nlines spectral lines prominent across most of our sample for which relatively uncontaminated continuum regions could be defined. These features and the continuum regions, one on either side of each feature, are listed in Table \ref{Tab:lines}.  To measure the EW of a feature, we first mitigated the effect of finite pixel sizes by linearly interpolating each spectrum onto a ten-times oversampled wavelength grid with uniform spacing in wavelength. The continuum was estimated by linear interpolation between the median fluxes of the
two continuum regions. We then applied the trapezoidal rule to numerically integrate the flux within the feature. We also measured ten spectral indices. We considered three indices quantifying the deformation in the continuum due to water absorption: the $\hind$ index, introduced in \S\ref{Sec:errors} (R12), the $\mathrm{H}_2\mathrm{O\mbox{-}H}$ index \citep{Terrien2012} and the $\mathrm{H}_2\mathrm{O\mbox{-}J}$ index \citep{Mann2013}. We also measured the flux ratios defined by \citet[]{McLean2003} and used by \citet{Cushing2005}. These ratios quantify absorption in several water, FeH and CO bands. The indices we measured are summarized in Table \ref{Tab:indices}. Finally, we considered three non-linear combinations of parameters. The non-linear combinations we considered were motivated by previous work: \citet[]{Luhman1999} suggested that $\naco$ is temperature-sensitive and R12 used the ratios $\nahind$ and $\cahind$ to fit their metallicity relation. 

\begin{deluxetable*}{l r r r r r r l}
\tablecaption{\label{Tab:lines}Spectral features searched as part of metallicity calibration}
\tablecolumns{8}
\tablehead{\colhead{Name} & \multicolumn{2}{c}{Feature} & \multicolumn{2}{c}{Blue continuum} & \multicolumn{2}{c}{Red continuum}	& \colhead{Source} \\ \colhead{} & \multicolumn{2}{c}{($\micron$)} & \multicolumn{2}{c}{($\micron$)} & \multicolumn{2}{c}{($\micron$)} & \colhead{}}
\startdata
\ion{Na}{1}&0.8180&0.8205&0.8140&0.8170&0.8235&0.8290& \citet{Cushing2005}\tablenotemark{a}\\
FeH&0.9895&0.9943&0.9850&0.9890&1.0150&1.0210& \citet{Cushing2005}\\
\ion{Na}{1}&1.1370&1.1415&1.1270&1.1320&1.1460&1.1580& \citet{Cushing2005}\\
\ion{K}{1}, \ion{Fe}{1}&1.1682&1.1700&1.1650&1.1678&1.1710&1.1750& \citet{Cushing2005}\\
\ion{K}{1}, \ion{Fe}{1}&1.1765&1.1792&1.1710&1.1750&1.1910&1.1930& \citet{Cushing2005}\\
\ion{Mg}{1}&1.1820&1.1840&1.1710&1.1750&1.1910&1.1930& \citet{Cushing2005}\\
\ion{Fe}{1}&1.1880&1.1900&1.1710&1.1750&1.1910&1.1930& \citet{Cushing2005}\\
\ion{Fe}{1}&1.1970&1.1985&1.1945&1.1970&1.1990&1.2130& \citet{Cushing2005}\\
\ion{K}{1}&1.2425&1.2450&1.2300&1.2380&1.2550&1.2600& \citet{Cushing2005}\\
\ion{K}{1}&1.2518&1.2538&1.2300&1.2380&1.2550&1.2600& \citet{Cushing2005}\\
\ion{Al}{1}&1.3115&1.3165&1.3050&1.3110&1.3200&1.3250& \citet{Cushing2005}\\
\ion{Mg}{1}&1.4872&1.4892&1.4790&1.4850&1.4900&1.4950& \citet{Cushing2005}\\
\ion{Mg}{1}&1.5020&1.5060&1.4957&1.5002&1.5072&1.5117& \citet{Covey2010}\\
\ion{K}{1}&1.5152&1.5192&1.5085&1.5125&1.5210&1.5250& \citet{Covey2010}\\
\ion{Mg}{1}&1.5740&1.5780&1.5640&1.5690&1.5785&1.5815&  \citet{Cushing2005}\\
\ion{Si}{1}&1.5875&1.5925&1.5845&1.5875&1.5925&1.5955& \citet{Covey2010}\\
CO&1.6190&1.6220&1.6120&1.6150&1.6265&1.6295& \citet{Covey2010}\tablenotemark{b}\\
\ion{Al}{1}&1.6700&1.6775&1.6550&1.6650&1.6780&1.6820&  \citet{Cushing2005}\\
Feature\tablenotemark{c}&1.7060&1.7090&1.7025&1.7055&1.7130&1.7160& \citet{Covey2010}\\
\ion{Mg}{1}&1.7095&1.7130&1.7025&1.7055&1.7130&1.7160& \citet{Covey2010}\tablenotemark{b}\\
\ion{Ca}{1}&1.9442&1.9526&1.9350&1.9420&1.9651&1.9701&  \citet{Cushing2005}\\
\ion{Ca}{1}&1.9755&1.9885&1.9651&1.9701&1.9952&2.0003& \citet{Covey2010}\\
Br-$\gamma$&2.1650&2.1675&2.1550&2.1600&2.1710&2.1740&  \citet{Cushing2005}\\
\ion{Na}{1}&2.2040&2.2110&2.1930&2.1970&2.2140&2.2200& \citet{Covey2010}\\
\ion{Ca}{1}&2.2605&2.2675&2.2557&2.2603&2.2678&2.2722& \citet{Covey2010}\\
CO&2.2925&2.3150&2.2845&2.2915&2.3165&2.3205& \citet{Covey2010}\\
CO&2.3440&2.3470&2.3410&2.3440&2.3475&2.3505& \citet{Covey2010}\\
\enddata 
\tablenotetext{a}{Atomic features were identified in \citet{Cushing2005}, but feature and continuum windows were defined based on our observations.}
\tablenotetext{b}{Feature and continuum windows were modified from those defined in \citet{Covey2010}.}
\tablenotetext{c}{Atomic feature not identified.}
\end{deluxetable*}
\begin{deluxetable*}{l l l l}
\tablecaption{\label{Tab:indices}Spectral indices searched as part of metallicity calibration}
\tablecolumns{4}
\tablehead{\colhead{Name} & \colhead{Absorption band} & \colhead{Definition} & \colhead{Source} }
\startdata
$\water\mbox{-}J$ & \J-band water deformation & $\frac{\la1.210-1.230\ra/\la1.313-1.333\ra}{\la1.313-1.333\ra/\la1.331-1.351\ra} $ & \citet{Mann2013} \\
$\water\mbox{-}H$ & \H-band water deformation & $\frac{\la1.595-1.615\ra/\la1.680-1.700\ra}{\la1.680-1.700\ra/\la1.760-1.780\ra} $ & \citet{Terrien2012} \\
$\hind$ & \K-band water deformation & $\frac{\la2.070-2.090\ra/\la2.235-2.255\ra}{\la2.235-2.255\ra/\la2.360-2.380\ra} $ & \citet{Rojas-Ayala2012} \\
$\water\mathrm{A}$ & $1.35\micron$ $\water$ band & $\langle 1.341-1.345\rangle /\langle 1.311-1.315\rangle$ & \citet{McLean2003} \\
$\water\mathrm{B}$ & $1.4\micron$ $\water$ band & $\langle 1.454-1.458\rangle /\langle 1.568-1.472\rangle$ & \citet{McLean2003} \\
$\water\mathrm{C}$ & $1.7\micron$ $\water$ band & $\langle 1.786-1.790\rangle /\langle 1.720-1.724\rangle$ & \citet{McLean2003} \\
$\water\mathrm{D}$ & $2.0\micron$ $\water$ band & $\langle 1.962-1.966\rangle /\langle 2.073-2.077\rangle$ & \citet{McLean2003} \\
CO & $2.29\micron$ $^{12}\mathrm{CO}$ 2-0 band & $\langle 2.298-2.302\rangle /\langle 2.283-2.287\rangle$ & \citet{McLean2003} \\
\J-FeH & $1.17\micron$ FeH 0-1 band & $\langle 1.198-1.202\rangle /\langle 1.183-1.187\rangle$ & \citet{McLean2003} \\
\Z-FeH & $0.99\micron$ FeH 0-0 band & $\langle 0.990-0.994\rangle /\langle 0.984-0.988\rangle$ & \citet{McLean2003} \\
\enddata
\tablecomments{Angle brackets denote the median of the wavelength range indicated. All wavelengths are in microns.}
\end{deluxetable*}


We searched for the combination of three parameters that provide the best fit to metallicity, using the forms:
\begin{align}
\feh&=A\left(\mathrm{F}_1\right)+B\left(\mathrm{F}_2\right)+C\left(\mathrm{F}_3\right)+D \\
&=A\left(\mathrm{F}_1\right)+B\left(\mathrm{F}_1\right)^2+C\left(\mathrm{F}_2\right)+D \\
&=A\left(\mathrm{F}_1\right)+B\left(\mathrm{F}_1\right)^2+C\left(\mathrm{F}_1\right)^3+D 
\end{align} 
where $F_n$ is the EW of one of the \nlines spectral features in Table \ref{Tab:lines}, one of the ten indices in Table \ref{Tab:indices}, or one of the three non-linear combinations of parameters described above. We used the $\rmse$ as a diagnostic to identify the best potential metallicity relations.

\begin{figure}
\includegraphics[width=\linewidth]{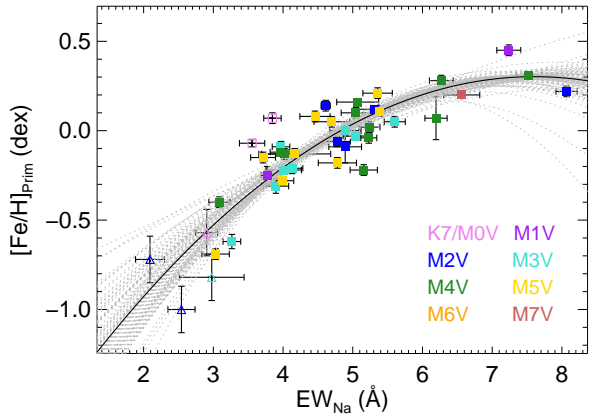}
\caption{Our best-fitting empirical metallicity relation (solid black line). We used a quadratic to relate the EW of the $\na$ line at $2.2\micron$ to the $\feh$ of an M dwarf. Our relation was calibrated against 36 M dwarfs in wide binaries with an FGK star of known metallicity. The NIR spectral type of each star is indicated by its color. The two K7/M0V stars that were not included in the calibration sample are plotted as open squares. We show an additional four M dwarfs for which the primary star has a metallicity measurement from \citet{Sozzetti2009} as open triangles; we used these stars to validate extrapolation of our relation to lower metallicities. Also shown (as dashed grey lines) are the best fits for 100 bootstrap samples.
\label{Fig:bestfit}}
\end{figure}

There were a multitude of relations with $\rmse<0.14\dex$, of which the majority included the EW of the $\na$ line at $2.2\micron$ as the primary indicator of metallicity (sometimes appearing as $\naco$ or $\nahind$) and included a quadratic term. However, we preferred the two-parameter fit $\feh=A\left(\ewna\right)+B\left(\ewna\right)^2+C$ because it uses one fewer parameter. Motivated by the clear trend with metallicity seen in $\ewna$, we also considered functional forms other than a quadratic, including a spline. No other forms tested resulted in a statistically superior fit. We show our result in Figure \ref{Fig:bestfit}. 

In performing our final fit, we did not include the two K7/M0V stars. In Figure \ref{Fig:bestfit} these are evident as having an $\ewna$ lower than other calibrators of similar metallicity. As discussed in \S\ref{Sec:bias}, we attempted to find a metallicity relation that was valid through these early spectral types, but did not converge on a suitable result.  Our final calibration sample therefore includes 36 M dwarfs with spectral types M1V and later. We address this choice in detail in the following section. 

Our best fit is given by:
\begin{align}\label{Eq:bestfit}
\feh = - 1.96\dex+ 0.596\dex\left(\mathrm{EW}_\na/\mathrm{\AA}\right) \\- 0.0392\dex\left(\mathrm{EW}_\na/\mathrm{\AA}\right)^2 
\end{align}
It is calibrated for $\ewna$ between $3$ and $7.5$\AA, corresponding to metallicities of $-0.6<\feh<0.3\dex$, and for NIR spectral types from M1V to M5V. There are indications that $\ewna$ begins to saturate for $\feh>0.3 \dex$ and our best fit becomes multivalued for $\ewna>7.5$\AA, so the calibration cannot be extrapolated past this point. 
The four M dwarfs for which the primary star has a metallicity measured by  \citet[]{Sozzetti2009}  objects indicate that our relation appears to be valid when extrapolated to $\ewna=2$\AA, corresponding to $\feh=-1.0\dex$.  
In \S\ref{Sec:met-check} we confirm the validity of the relation for later NIR spectral types by comparing metallicities
estimated for members of CPM M-dwarf multiples with a range of spectral types.  While there is only one calibrator later than M5, this object also indicates that the relation can be extrapolated as late as M7.

We estimated the error introduced by our limited number of calibrators by bootstrapping. We randomly selected 36 of our calibrators, allowing repeats, and re-fit our metallicity relation. The standard deviation of the difference between the best fitting metallicities of the M dwarf secondaries and the metallicities of the primaries, averaged over 100 bootstrap samples, was $0.12\pm0.01 \dex$. The correlation coefficient, $\Rap$ is often used to evaluate the goodness of fit. The correlation coefficient indicates how well the fit explains the variance present in the data and is given by:
\begin{equation}
\Rap=1-\frac{(n-1)\sum(y_{i,model}-y_i)^2}{(n-p)\sum(y_i-\bar{y})^2}
\end{equation}
where $n$ is the number of data points and $p$ is the number of parameters. The $\Rap$ value for our fit is $0.78\pm0.07$.
The best-fitting metallicities for our calibrators are included in Table \ref{Tab:binaries-measurements}. The errors on metallicity include the errors on $\ewna$, bootstrap errors and the scatter in our best fit, added in quadrature. We took the bootstrap errors to be the $1\sigma$ confidence interval on the resulting metallicities when considering the best fits from 100 bootstrap samples. The intrinsic scatter in the relation ($0.12\dex$) dominates for all but the lowest metallicity stars.

The scatter in our metallicity relation is similar to those reported by R10, R12, \citet[]{Terrien2012} and \citet{Mann2013} despite differences in sample size, lending support to the idea that the scatter is astrophysical in origin. We consider potential temperature and surface gravity effects in \S\ref{Sec:bias}. One possibility is variations between the $\na$ abundance and $\feh$ of the primary solar-type star. We considered whether an M dwarf's $\ewna$ is a better tracer of its primary star's $\na$ abundance than its $\mathrm{Fe}$ abundance. $32$ of our calibrators have measured abundances for $\na$ from \spocs. We related the spectral features and indices in Tables \ref{Tab:lines} and \ref{Tab:indices} to the $\na$ abundance of the primary star.  We found several suitable tracers; however, none reduced the scatter.

In Table A1, we include the EWs of the Na line at $2.20\micron$ and the Ca line at $2.26\micron$, the $\hind$ index, our inferred $\feh$, and their associated errors for each of our targets. The corresponding values for the FGK-M CPM pairs can be found in Table \ref{Tab:binaries-measurements}.

\subsection{Influence of effective temperature and surface gravity on the metallicity calibration}\label{Sec:bias}
 
\begin{figure}
\includegraphics[width=\linewidth]{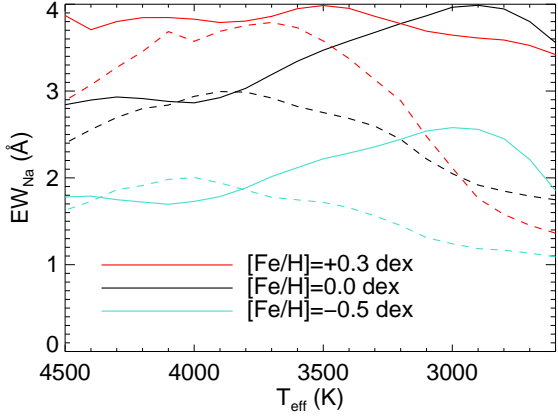}
\caption{The behavior of the $\na$ line at $2.2\micron$ in the BT-Settl stellar models \cite{AllardF.2011}. The horizontal axis is model effective temperature, approximately corresponding the spectral type range K5V-M6.5V. The vertical axis shows measure $\ewna$ in \AA. Dashed lines indicate theoretical spectra with $\log g=4$ and solid lines those with $\log g=5$.
\label{Fig:nabt}}
\end{figure}

We examine the potential influence of differences in the effective temperature and surface gravity on the metallicity calibration presented in \S\ref{Sec:relation} by computing $\ewna$ for a grid of BT-Settl theoretical spectra for spectral types K5V-M7V, shown in Figure \ref{Fig:nabt} \citep[][the behavior of NIR lines in theoretical spectra are discussed in some detail in R12]{AllardF.2011}. The spectral type range corresponds to approximately K5V-M6.5V, depending on the adopted temperature scale (we quote the temperature scale from E. Mamajek, which is available online.\footnote{http://www.pas.rochester.edu/\~emamajek/EEM\_dwarf\_UBVIJHK\_colors\_Teff.dat}). The BT-Settl theoretical spectra show $\ewna$ varying by $1$\AA\ between M0V and M8V stars (Figure \ref{Fig:nabt}). We also note that in our \K-band \spex\ spectral sequence (Figure \ref{Fig:masterK}) the $\na$ line at $2.2\micron$ is broader for the latest spectral types. 

\begin{figure}
\includegraphics[width=\linewidth]{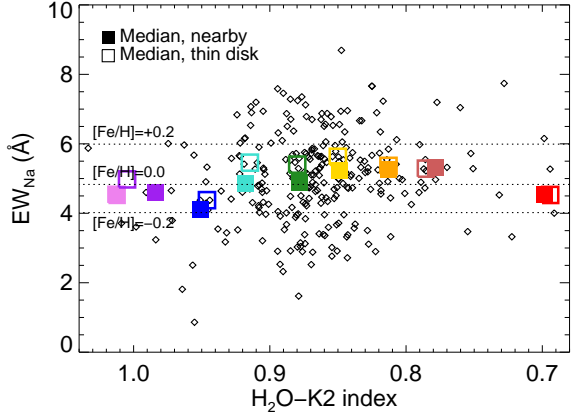}
\caption{The behavior of $\ewna$ in our observed spectra. We plot the median $\ewna$ against the median $\hind$ for each NIR spectral type as shown in Figure \ref{Fig:bestfit}. The medians for two subsamples are shown. Filled squares include only those stars which are in our nearby sample and open squares include only kinematically young stars. Points are colored by their NIR spectral type, from purple for M0V stars to red for M8V stars, as shown in Figure 11. \label{Fig:na}}
\end{figure}

We plot in Figure \ref{Fig:na} the median $\ewna$ for each NIR spectral type as a function of $\hind$, for two subsamples. Our ``nearby sample'' (\S\ref{Sec:spectroscopy}) formed the first, and kinematically young stars ($\Vs<50\mathrm{km/s}$) formed the second. We selected the nearby sample to approximate a volume limited sample, which is unlikely to be influenced by selection effects that may exist in the rotation sample. We selected the kinematically young sample in order to isolate stars that are expected to be of similar age and metallicity. We found a similar increase in the median $\ewna$ of mid to late M dwarfs as we noted in the theoretical spectra. This could introduce a systematic error of $0.1\dex$ in the metallicities of early M dwarfs relative to mid M dwarfs. However, we are uncertain of the origin of this effect, given the differing behavior of our two subsamples and the relative differences in the number of early and late type stars (there are 23 stars with NIR spectral types M0V-M2V and 231 with spectral types M4V-M5V across the two subsamples).


\begin{figure}
\includegraphics[width=\linewidth]{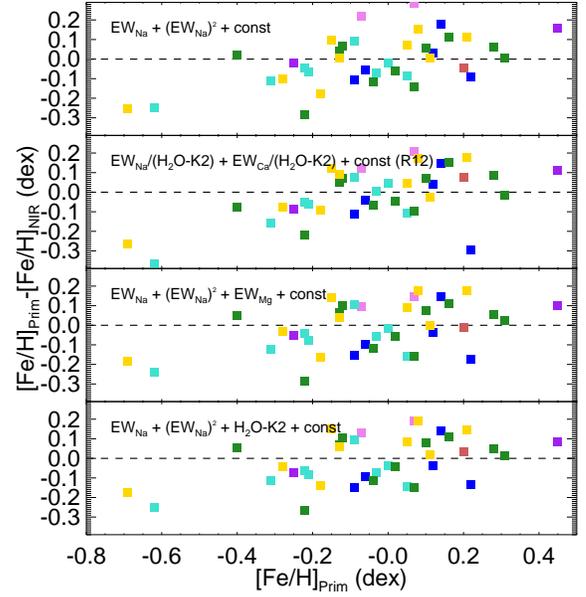}
\caption{The residuals for the best-fitting metallicity relations for four different parameterizations. We include the K7/M0V calibrators in this analysis. Points are colored by their NIR spectral type, from purple for M0V stars to light red for M7V stars, as shown in Figure \ref{Fig:bestfit}.
\label{Fig:twoparam}}
\end{figure}

\begin{figure}
\includegraphics[width=\linewidth]{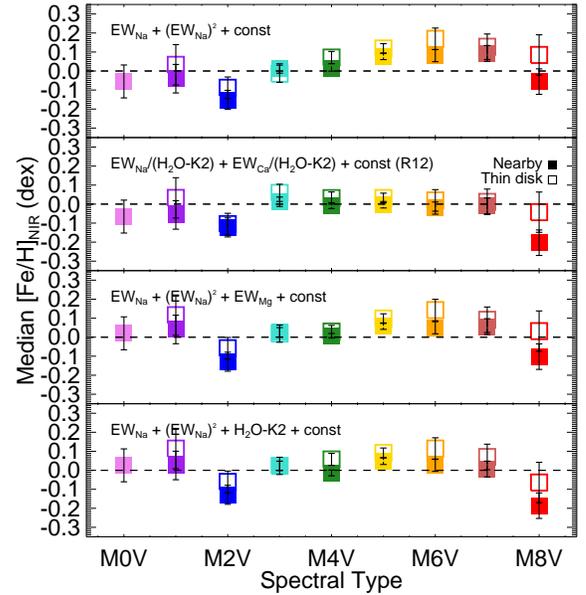}
\caption{The median metallicity for two subsamples of stars as a function of NIR spectral type. Filled squares indicate median metallicities for stars without measured rotation periods and open squares indicate the median metallicities for kinematically young stars. Points are colored by their NIR spectral type, from purple for M0V stars to red for M8V stars, as shown in Figure \ref{Fig:bestfit}.
\label{Fig:twoparam_sp}}
\end{figure}

We considered whether an alternative parameterization could account for this potential bias. We show the residuals for our chosen parameterization and three alternatives, including the parameterization from R12, in Figure \ref{Fig:twoparam}. In Figure \ref{Fig:twoparam_sp}, we show the effect that the alternative calibrations have on the metallicities of the sample as a whole. With the R12 parameterization, the inferred metallicities of the latest stars decreased by $0.1\dex$ and metallicities were consistent across spectral types. However, the metallicities of M5 were lowered relative to those of M4 dwarfs, the spectral range across which our relation is best calibrated. Furthermore, the fit is unconstrained at the latest spectral types where the choice of the R12 parameterization makes the most difference. Including the EW of magnesium or the $\hind$ as a third parameter in the metallicity calibration improves the fit for the two K7/M0V calibrators and has only a marginal effect at other spectral types. However, only scatter \emph{above} the best fit plotted in Figure \ref{Fig:bestfit} was reduced in this case, while the scatter \emph{below} our best fit remained. 

When the M0V calibrators were not included in the fit, the
addition of these extra parameters makes little difference.
Therefore, rather than including an additional parameter to fit these
two points at the far end of our spectral type range, we simply limit
our calibration to a range of spectral types which appear to be
well-fit by a relation depending solely on $\na$.


The insensitivity of NIR spectral types to late K dwarfs may be partially responsible for the behavior seen in our two M0V calibrators. The optical spectral type of PM I07400$-$0336 places it as K6.5V dwarf \citep[]{Poveda2009} and LSPM J1030+5559 has been identified previously as a K7V dwarf \citep[]{GarciaB.1989}. However, theoretical models indicate that the $\ewna$ should remain constant between late M and mid K dwarfs (with slight dependence on surface gravity), and \citet[]{Mann2013} reported a metallicity calibration that is valid from K5V-M5V. 

\begin{figure}
\includegraphics[width=\linewidth]{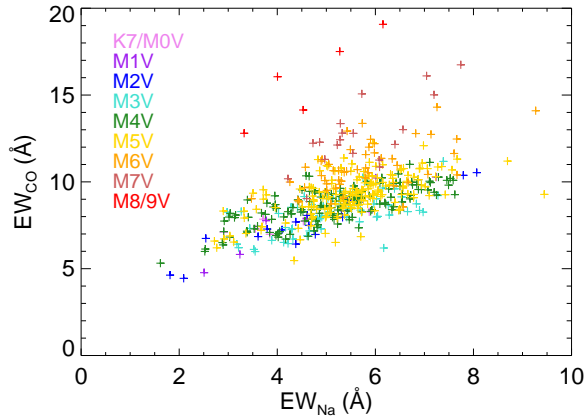}
\caption{We compare $\ewna$ against $\mathrm{EW}_\co$ for all stars in our sample. According to \citet[]{Luhman1998}, very young stars would reveal themselves through low $\ewna$ but high $\mathrm{EW}_\co$. We have no data in the upper left corner of this plot, indicating it is likely that no very young stars are present in our data.
\label{Fig:naco}}
\end{figure}

Surface gravity remains one possible explanation for the K7/M0V discrepancy and has yet to be explored in the context of empirical calibrations. \citet[]{Luhman1998} demonstrated that in the low surface gravity environments of very young stars, $\na$ may appear abnormally weak. It is therefore possible that an M dwarf with an age of several Myr could be masquerading as a metal-poor object. The $\co$\ band head is sensitive to gravity in the opposite manner and is therefore a useful indicator of youth \citep[]{Luhman1998}. In Figure \ref{Fig:naco}, we plot $\ewna$ against $\mathrm{EW}_\mathrm{CO}$ for all stars in our sample. We found a general positive correlation and spectral dependence, but no object stood apart has having low $\ewna$ but high $\mathrm{EW}_\co$. This is not surprising as it is unlikely that we would find a new, bright young star within $25$pc.

We considered the potential for other systematics by comparing the difference between our best fitting metallicities and the metallicities of the primaries to the EWs of all other indices. In all cases, we found no significant systematic effects.

\subsection{Tests of our metallicity relation}\label{Sec:met-check} 

\begin{figure}
\includegraphics[width=\linewidth]{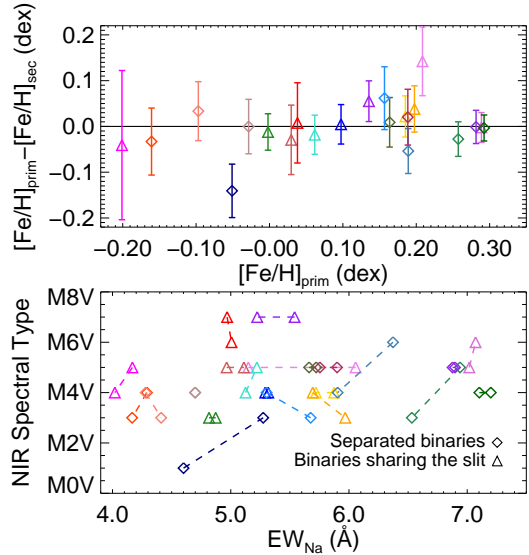}
\caption{We compare measurements of M dwarf-M dwarf CPM pairs. In the top panel, we plot the $\feh$\ difference against the metallicity of the earlier M dwarf in the pair. The mean $\feh$ difference between pairs is $-0.01$ dex and the standard deviation is $0.05$ dex. In the bottom panel, we compare $\ewna$ measurements and spectral types of the binaries. Points are color-coded such that a pair has the same color in the top and bottom panels.
\label{Fig:binaries}}
\end{figure}

As a test of our metallicity calibration, we compared the metallicities we estimated for the components of M dwarf-M dwarf CPM pairs. We have observed 22 such pairs. 11 were placed on the slit together and so share observing conditions, while 11 were observed separately but close in time. In both cases, the two stars were reduced with the same telluric standard. In Figure \ref{Fig:binaries} we show the results of this comparison. The mean metallicity difference between the primary and secondary components is $-0.01\dex$ with a standard deviation of $0.05\dex$. This is less than the uncertainty of our metallicity measurement by a significant amount, lending support to the idea that most of the scatter in the metallicity relation is astrophysical in origin, as mentioned in \S\ref{Sec:relation}.

We also compared $\ewna$ measurements for stars that were observed on more than one occasion in Figure \ref{Fig:multobs-line} (see \S\ref{Sec:errors}). We found that our $\ewna$ measurements were consistent even for observations taken in very different conditions and separated in time by months or more. The mean $\ewna$ difference between the observation we elected to keep and the observation we discarded was $-0.01\dex$ with a standard deviation of $0.04\dex$.

\subsection{Inclusion of previous metallicity estimates}\label{Sec:tspec}

\begin{figure}
\includegraphics[width=\linewidth]{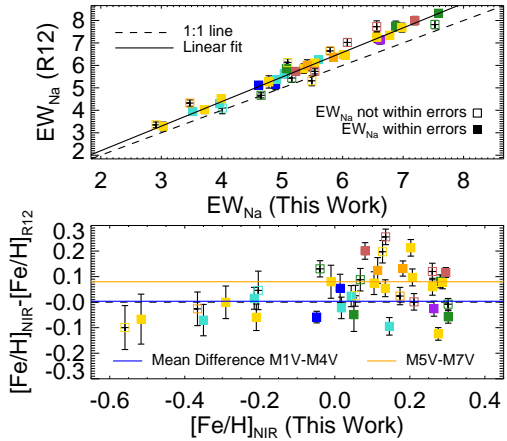}
\caption{Comparison between our measurements and those from R12. In the top panel, we compare $\ewna$ measured in this work using the \spex\ instrument on IRTF to those presented in R12, who used the TripleSpec instrument on Palomar. We show the one-to-one line (dashed line) and our best fit (solid line). In the bottom panel, we compare $\feh$ estimated in this work directly to that estimated by R12. We over plot the mean metallicity difference for an early subsample (NIR spectral types M1V-M4V) and a late subsample (M5V-M7V). Data are plotted as filled squares if our $\ewna$ measurements agree within the errors and as open squares if the discrepancy is larger than the associated error. In both panels, data are colored by their NIR spectral type, from purple for M0V stars to red for M8V stars, as shown in Figure \ref{Fig:bestfit}. \label{Fig:tspec}}
\end{figure}

R12 published their measurements of $\ewna$, $\mathrm{EW}_\mathrm{Ca}$ and $\feh$ for $133$ M dwarfs using the TripleSpec instrument on Palomar \citep{Herter2008}. To facilitate joint use of our observations and those from R12, we determined the relationship between TripleSpec and \spex\ EWs. We compare our $\ewna$ measurements directly in Figure \ref{Fig:tspec}. We used the following relation to convert from TripleSpec to IRTF $\ewna$:
\begin{equation}\label{Eq:na}
\mathrm{EW}_\mathrm{Na,N13}=0.036+0.90\left(\mathrm{EW}_\mathrm{Na,R12}\right)
\end{equation}
Similarly for the Ca line at $2.26\micron$:
\begin{equation}\label{Eq:ca}
\mathrm{EW}_\mathrm{Ca,N13}=0.22+0.88\left(\mathrm{EW}_\mathrm{Ca,R12}\right)
\end{equation}

We also directly compared our metallicity estimates for the 28 stars in common (excluding metallicity calibrators). As seen in Figure \ref{Fig:tspec}, the metallicity measurements agreed well for sub-solar metallicities, but for metallicities above solar, the relation in this work gives higher metallicities for late M dwarfs (M5V-M7V). The difference between our inferred metallicity and that from R12 is $0.0\pm0.07\dex$ for M1V-M4V stars and $0.08\pm0.05$ for M5V-M7V stars. This difference is consistent with the effects discussed in \S\ref{Sec:bias}, but we note that our relation is most strongly constrained for M4 and M5 dwarfs. 

The objects observed by R12 are listed in Table A2. We have included EWs updated using Equations \ref{Eq:na} and \ref{Eq:ca}. After applying our $\ewna$ relationship, we can directly compute the metallicities for stars published in R12 using our  metallicity calibration. We also present these updated metallicities in Table A2.

\section{Photometric metallicity calibrations}\label{Sec:colors}

\begin{figure*}
\includegraphics[width=\linewidth]{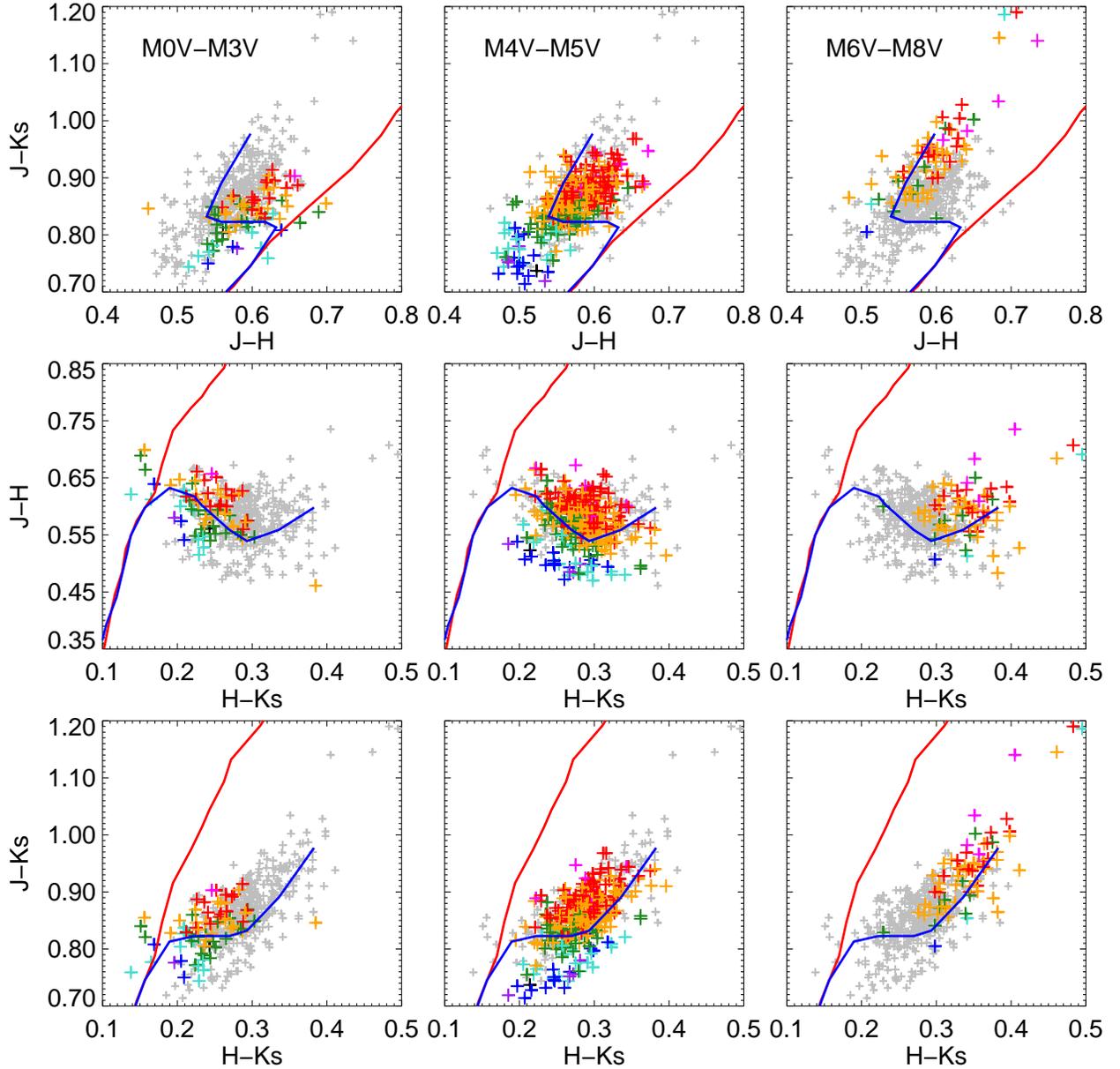}
\caption{Color-color diagrams for M dwarfs observed with IRTF. Stars are colored by the metallicity we estimated from the NIR. Stars with $\ewna<2\mathrm{\AA}$ are plotted in black. Those with $-1.0<\feh<-0.6$ are in purple, with $-0.6<\feh<-0.4$ in blue, with $-0.4<\feh<-0.2$ in cyan, with $-0.2<\feh<0.0$ in green, with $0.0<\feh<+0.2$ in orange, and with $+0.2<\feh<+0.3$ in red. Stars with $\ewna>7.5$\AA\ are plotted in magenta. Grey points are stars of other spectral types other than the range indicated in the top panels. Overplotted are the dwarf (blue) and giant (red) tracks from \citet[]{Bessell1988}, converted to the 2MASS system using the updated color transformations of \citet{Carpenter2001}, which are available online. \label{Fig:color}}
\end{figure*}

We exploited our sample of M dwarfs with spectroscopically determined NIR metallicities to identify which color-color diagrams are metallicity sensitive and to derive an empirical relationship between an M dwarf's NIR color and its metallicity. In Figure \ref{Fig:color}, we plot $JHK_S$ color-color diagrams for the $444$ of our targets with the highest quality 2MASS magnitudes \citep[\texttt{qual\_flag}=AAA]{Skrutskie2006}. We also plot the \citet{Bessell1988} M dwarf main sequence, which coincides with our solar metallicity stars. These diagrams are plotted in the 2MASS photometric system; we used the color transformations updated\footnote{http://www.astro.caltech.edu/\~jmc/2mass/v3/transformations/} from \citet{Carpenter2001} to transform the colors from \citet{Bessell1988} to the 2MASS system.

\begin{figure}
\includegraphics[width=\linewidth]{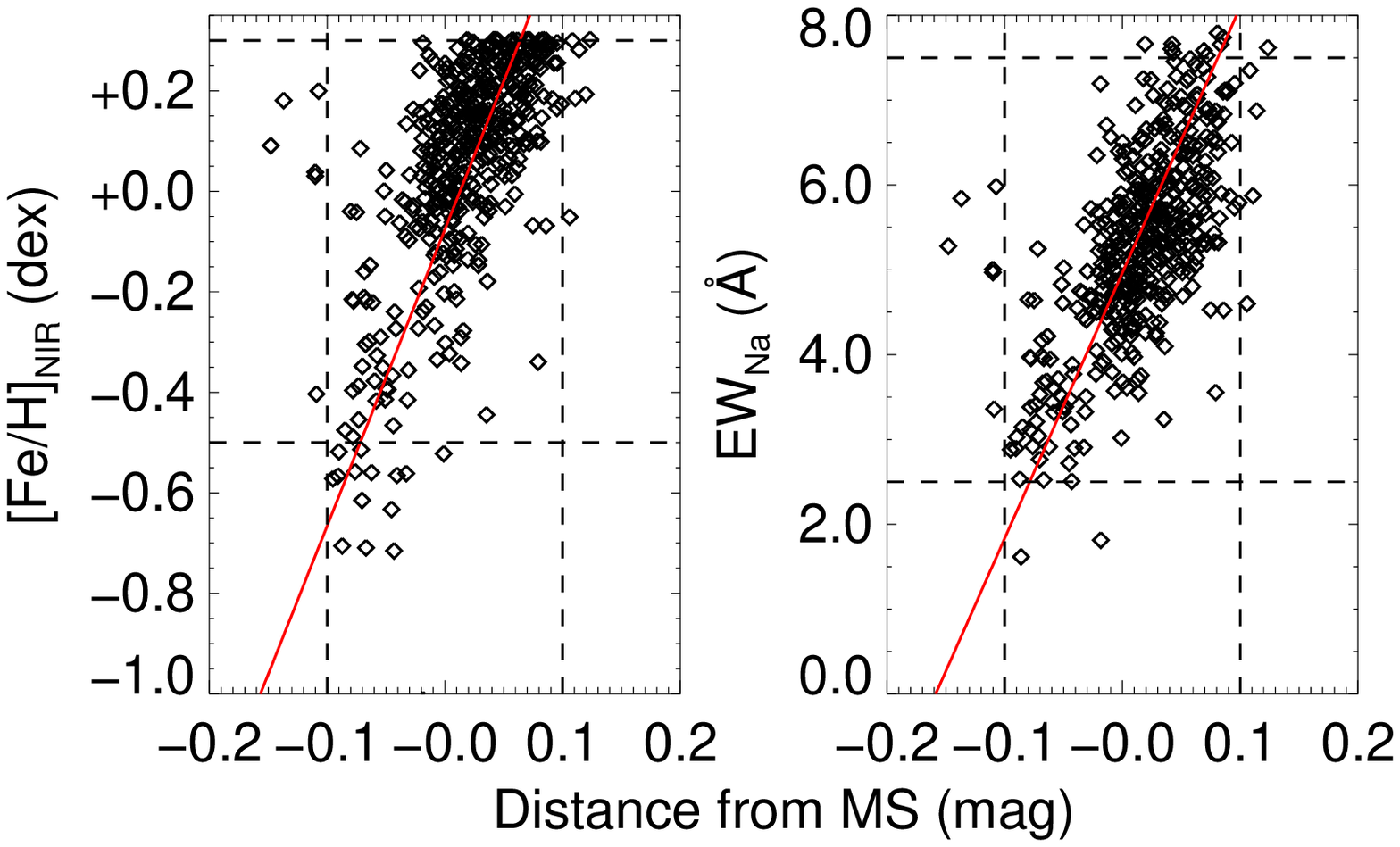}
\caption{Metallicity (as measured from the NIR; left) and $\ewna$ (right) plotted against distance from the Bessell \& Brett main sequence. Our best-fit calibration for an M dwarf's metallicity or $\ewna$ as a function of the distance from the main sequence is over plotted in red. The range over which the calibration is valid is included as dashed vertical lines. 
\label{Fig:dist}}
\end{figure}

All color combinations discriminated effectively between low and high metallicity stars. Consistent with \citet[]{Johnson2012}, we found that the $J-K_S$ color of an M dwarf is the best single-color diagnostic of its metallicity. We used the vertical ($J-K_S$) distance from the $J-K_S$, $H-K_S$ Bessell \& Brett dwarf main sequence ($\dms$) as the diagnostic for the metallicity of an M dwarf. We considered using $\dms$ to determine both $\ewna$ and $\feh$ directly (Figure \ref{Fig:dist}). We chose to relate $\dms$ to $\ewna$ because the correspondence is linear and because it relates two directly measured quantities. 

\begin{figure}
\includegraphics[width=\linewidth]{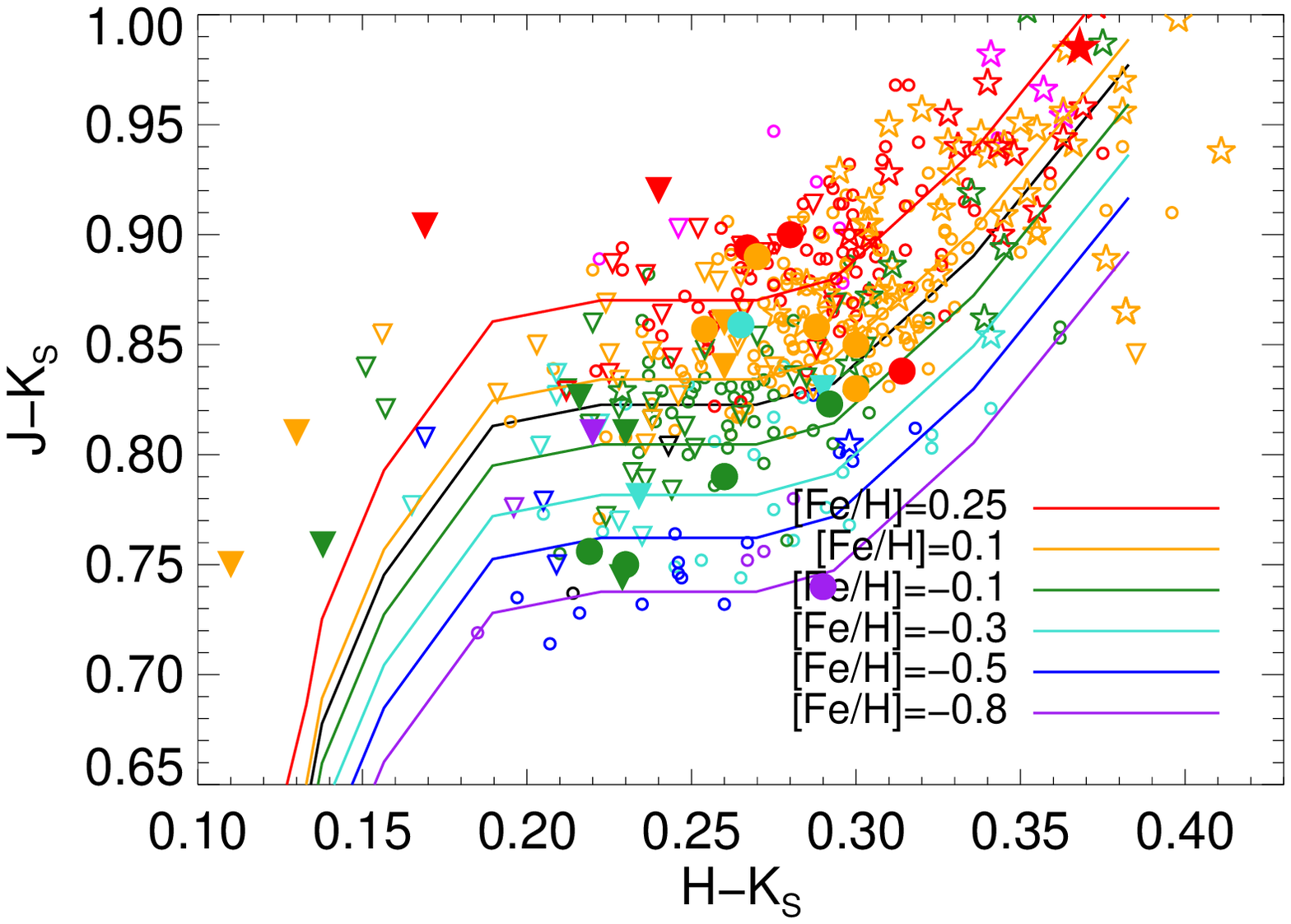}
\caption{Reproduction of the $J-K_S$, $H-K_S$ color-color diagram for all M dwarfs observed with IRTF. Stars are colored as in Figure \ref{Fig:color}, while symbols indicate NIR spectral type (K7V-M3V as triangles, M4V-M5V as circles, and M6V-M9V as stars). Large filled symbols are our metallicity calibrators. Overplotted are isometallicity contours for our best fit, which relate distance from the main sequence to metallicity via $\ewna$.
\label{Fig:contours}}
\end{figure}

We determined the relation between $\ewna$ and $\dms$ using those stars with $2.5<\ewna(\mathrm{\AA})<7.5$ and $|\dms|<0.1$. We binned the data into $0.5$\AA-wide bins and computed the median $\dms$ in each. We then fit a straight line through these points, using the reciprocal square root of the number of data points in each bin as the weights. The best-fitting relation between $\ewna$ and $\dms$, shown in Figure \ref{Fig:dist} is:
\begin{equation}
\ewna=4.97\mathrm{\AA} + 31.3\mathrm{\AA}\left(\dms/\mathrm{mag}\right)
\end{equation}
The standard deviation is $2.0$\AA\ and the $\Rap$ value is 0.92. We applied Equation \ref{Eq:bestfit} in order to write metallicity as a function of $\dms$:
\begin{align}
\feh=0.0299\dex + 6.47\dex\left(\dms/\mathrm{mag}\right) \\- 38.4\dex\left(\dms/\mathrm{mag}\right)^2
\end{align}
We show the resulting photometric metallicity calibration in Figure \ref{Fig:contours}.

Our calibration is valid from $2.5<\ewna(\mathrm{\AA})<7.5$, corresponding to $-0.7<\feh<0.3$  and for $0.2<H-K_S<0.35$. The $1\sigma$ uncertainty in $\ewna$ translates to $0.1$ dex for $\ewna=7$\AA\ and $0.5$ dex for $\ewna=3$\AA. This calibration is particularly useful because it does not require \V magnitudes, which are often unreliable, or parallaxes, which are often unavailable. In contrast, accurate $JHK_S$ magnitudes are available for the majority of nearby M dwarfs from 2MASS.



\section{Radial velocities from NIR spectroscopy}\label{Sec:rv}

Absolute wavelength calibration for moderate resolution NIR spectra are typically determined using a lamp spectrum taken at the same pointing as the science spectrum, as done by \citet{Burgasser2007}, who measured the radial velocity of an L dwarf binary to $18 \kms$ accuracy using \spex ($R\approx2000$). An alternative is to take deep sky exposures and use OH emission lines to perform wavelength calibration. This approach was used, for example, by \citet{Muirhead2013}, who use the TripleSpec instrument on Palomar ($R\approx2700$) to measure absolute radial velocities for the eclipsing post common envelope binary KOI-256 with typical errors of $4 \kms$.  

We acquired Thorium-Argon spectra regularly throughout the night to track instrumental variations, but it was not possible to obtain them at every telescope position due to the exposure times required.  We found that this procedure was not adequate for accurate radial velocity work. We therefore used telluric absorption features to supplement the
wavelength calibration by adjusting the velocity zero-points for individual observations, then cross-correlated each spectrum with a standard spectrum to measure its absolute RV (\S\ref{Sec:rvmethod}). In \S\ref{Sec:chubak}, we discuss using precisely measured RVs from \citet{Chubak2012} to investigate random and systematic error. We describe further tests of our method in \S\ref{Sec:rv-check}. 

\subsection{Radial velocity method}\label{Sec:rvmethod}

Atmospheric absorption features present in our data provided a natural replacement to arc spectra. By correlating the telluric lines in our spectra with a theoretical atmospheric transmission spectrum (hereafter called simply the ``transmission spectrum''), we determined the absolute wavelength calibration. The \texttt{SpeXtool} package includes a transmission spectrum created using \texttt{ATRANS} \citep[]{Lord1992}. This spectrum was created using environmental parameters typical of Mauna Kea and an airmass of 1.2 and has a resolution five times that of \spex. We used the wavelength calibration determined by \spex\ using ThAr arc spectra as our initial wavelength guess for the nontelluric corrected science spectrum. From this wavelength solution, we created a wavelength vector that was oversampled by a factor of six and linearly spaced in wavelength. 

We found that excellent continuum removal was required for the wavelength calibration to be determined through direct cross correlation of the science spectrum and the transmission spectrum. However, the large atmospheric features made this difficult. Instead of attempting to remove the continuum from the M dwarf and subsequently finding the offset between the stellar spectrum and the atmospheric spectrum, we tackled these problems simultaneously. We did this by finding the modifications to the transmission spectra that provided the best match the telluric features observed in the science spectrum.  There were three differences between the theoretical transmission spectrum and the telluric features as observed in the science spectrum: the continuum, the strength of the telluric features and the pixel offset between the spectra.

The first parameter of our model was a Legendre polynomial as a function of pixel by which the transmission spectrum was multiplied in order to replicate the shape of the spectrum. The curvature of the spectrum was affected by both instrumental effects and the M dwarf spectral energy distribution. We used a 3rd or 4th degree Legendre polynomial and fit for the coefficients. We selected the degree of the polynomial by eye for each order, using the lowest degree polynomial required to model several representative M dwarf spectra. 

The second parameter was an exponential scaling of the flux, to account for the effects of airmass and atmospheric water vapor on the depths of telluric features. The transmission spectrum represents typical conditions on Mauna Kea, while we observed at air masses from $1.0$ to $1.7$ with humidity from $85\%$ to less than $15\%$. 

As discussed in \citet[]{Blake2010}, differences in airmass scale the depths of the telluric features ($T$) as $T=T_{0}^\tau$ where the optical depth $\tau$ scales linearly with airmass. \citet[]{Blake2010} were able to find a single linear scaling between airmass and $\tau$ using a large sample of A0V stars. We attempted to use the same approach, but found substantial scatter and systematic differences in the scaling of different telluric features with airmass.  This is likely due to the water absorption features in our spectral region, which are time-variable, and cannot be modeled by a simple function of airmass alone. We therefore chose to take an empirical approach and included the exponential scaling $\tau$ as a model parameter.

The third and final parameter was the offset in pixels between the transmission and science spectrum. We modeled the offset as linear in wavelength. To apply the shift, we created a new wavelength vector that was linearly shifted from the original and interpolated the transmission spectrum onto the new wavelength vector. 
We constrained the allowable range for the offset  because atmospheric features appear at regular spacing and we found that if unconstrained, our fitting program can too often land in a local minimum. We used $0.0008\micron$ as the limit, which is larger than any offset we expected. In our full sample, no shifts beyond $0.0006\micron$ were found, and very few beyond $0.0004\micron$.

We modeled each order of the non-telluric corrected science spectrum independently, minimizing the difference between our model and the science spectrum using a nonlinear least squares approach, implemented through \texttt{mpfit} \citep{Markwardt2009}. We determined by trial and error the region of each order to use. Regions with high signal to noise and strong telluric features but uncontaminated by strong stellar features were required for optimal performance. Because of these constraints, this method worked better in the \J, \H\ and \K-bands than it did in \Z or \I.

Once we determined the absolute wavelength solutions of science target and an RV standard, we interpolated the telluric-corrected spectra onto a common wavelength vector that was oversampled and uniform in the log of the wavelength (such that a radial velocity introduces a constant offset in pixels). The continuum is different in the telluric-corrected spectrum because telluric correction removed instrumental effects, so we used a spline to remove the continuum. We used \texttt{xcorl} to cross-correlate the two spectra and determine the offset. We used the same standard star (Luyten's star, also known as Gl 273 or LSPM J0727+0513) throughout because it had an accurately measured absolute radial velocity from \citet{Chubak2012} and a NIR spectral type in the middle of our range (M4V). 

We took the final RV for each target to be the median of the RVs measured in the \J, \H\ and \K-bands and applied the heliocentric correction, implemented through the IDL routine \texttt{baryvel} \citep{StumpffP.1980}. Our final estimate of the error is the $1\sigma$ confidence limit on the RV after 50 trials added in quadrature to $4.4\kms$ (our internal measurement error, see \S\ref{Sec:rv-check}). These values are reported in Table A1.

This method of measuring radial velocities is applicable to other
moderate resolution NIR spectrographs, including TripleSpec, and
uses observations of the target star to refine the wavelength
calibration. Our method is therefore likely to be useful for instruments where obtaining
lamp spectra is expensive.

\subsection{Using precise RVs to investigate errors and systematics}\label{Sec:chubak}

\begin{figure}
\includegraphics[width=\linewidth]{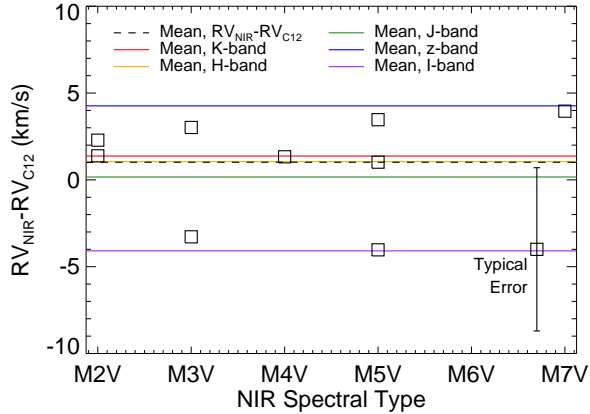}
\caption{We compare our RV measurements to those from \citet[]{Chubak2012}, with NIR spectral type on the horizontal axis. Data points show the difference between our adopted RV for each star, which is the median of the RV measured in each of the  \J, \H, and \K-bands, and that reported in \citet{Chubak2012}. The dashed line shows the mean difference between our measurements and those from \citet{Chubak2012}. We also look at how well the RV measured from a single band compares to the values from \citet{Chubak2012}; the mean difference for each band is plotted as a colored line. The \Z\ and \I -bands tend to over- and underestimate the RV. A $-2.6\kms$ offset has been applied.
\label{Fig:chubak}}
\end{figure}

\citet[]{Chubak2012} presented absolute, barycentric-corrected RVs for $2046$ dwarf stars with spectral types from F to M. M dwarf RVs were measured by comparison to an M3.5V RV standard, offset to agree with the measurements from \citet{Nidever2002}. No corrections were made for convective or gravitational effects for M dwarfs, and \citet{Chubak2012} report a systematic error of $0.3\kms$ (random errors are at this level or lower in nearly all cases). Ten of their M dwarfs are in our sample. We chose one of these, LSPM J0727+0513, as our standard star. For the other nine stars, we compare our measurements to those from \citet[]{Chubak2012} in Figure \ref{Fig:chubak}. Considering the RV measured in each order separately, we found that the bluest two bands (\I\ and \Z) systematically underestimate (\I-band) or overestimate (\Z-band) the RV. The wavelength calibration is also subject to failure in those bands. We suggest that this is because in these two orders, the strongest stellar features overlap with the strongest telluric features, compromising the wavelength calibration and therefore the velocity measurement. They were also the orders with the lowest S/N. The RVs reported in this paper are the median of the \J, \H, and \K-band measurements.

We measured RVs for all our targets using each of the ten RV standards from \citet{Chubak2012} in order to determine our internal error and systematic RV offset. The typical standard deviation of RVs measured against an alternative standard relative to that measured against LSPM J0727+0513 was $4.2\kms$. We used this value as our internal random error. RVs measured using LSPM J0727+0513 were systematically higher than those measured using other RV standards. Considering M3V-M5V standards, the median offset was $2.6\kms$ with a standard deviation of $1.5\kms$. The values reported in this paper include a $-2.6\kms$ systematic RV correction. Our total internal measurement error is $4.4\kms$, which is our internal random error ($4.2\kms$) added in quadrature to our internal systematic error ($1.5\kms$).

Our choice of a single, mid-M RV standard does not appear to
systematically affect the RV measurements or errors of early
and late M dwarfs at this level of precision. ÊWe investigated the
effect of the standard spectral type by comparing the results using
LSPM J0727+0513 with using an M2V star, PM I06523-0511 (Gl 250), to measure the RVs of
early M-dwarfs, and an M7V star, J1056+0700 (Gl 406), to measure the RVs of late
M-dwarfs, finding that these choices did not appear to systematically
affect the measured RVs, and that the scatter remained consistent with
our estimated uncertainties.

\subsection{Validating the use of \spex\ for radial velocities}\label{Sec:rv-check}

\begin{figure}
\includegraphics[width=\linewidth]{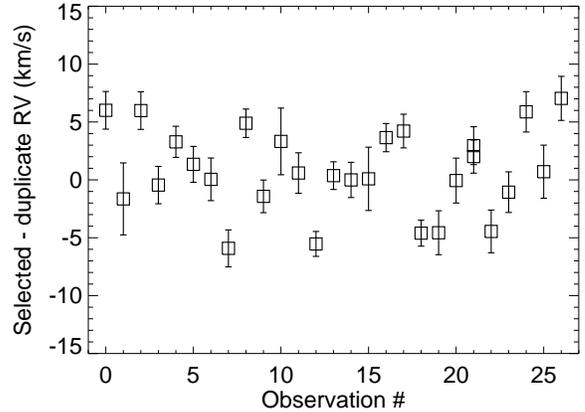}
\caption{We compare RV measurements for 26 stars which we observed multiple times. For each star, we plot the difference between the RV measured from the observation we elected to keep and the observation we did not use. The error bars plotted are the $1\sigma$ confidence intervals after 100 trials. 
\label{Fig:multobs-rv}}
\end{figure}

To determine the precision of our wavelength calibration method, we used the transmission spectrum to create simulated data in each order, which we then calibrated. We simulated stellar absorption lines of random widths, depths and locations on top of the transmission spectrum and multiplied by a polynomial (drawn from a random distribution) to curve the data. We then offset the spectrum and monitored how well we could recover that offset. The accuracy declined as more stellar absorption lines were added to the spectrum. With 50 added lines, accuracy was better than $5\kms$ in all orders and better than $1\kms$ in \H-band. 

We have multiple observations for 26 stars at different epochs. The time between observations ranges from days to months to years. We compared our RVs for these stars (Figure \ref{Fig:multobs-rv}). The mean difference between the observation we elected to keep and the observation we chose to discard with $0.08\kms$ with a  standard deviation of $4\kms$, consistent with our calculation of the error.

\begin{figure}
\includegraphics[width=\linewidth]{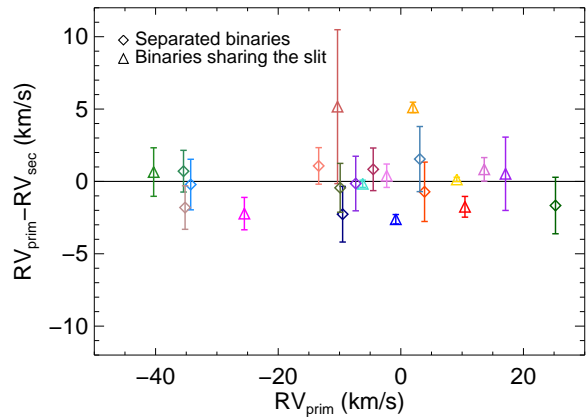}
\caption{We compare RV measurements for binary stars, 11 of which were observed independently and 11 of which were observed together on the slit. The error bars are the $1\sigma$ confidence limits in the RV after 100 trials. Colors uniquely identify pairs in this figure and in Figure \ref{Fig:binaries}.
\label{Fig:bin-rv}}
\end{figure}

Finally, we compared the RVs of CPM pairs (Figure \ref{Fig:bin-rv}). 11 of these stars are separated and were observed independently and 11 were observed together on the slit. These observations were taken close in time, at near-identical conditions and were reduced using the same wavelength calibration and telluric standard. The mean RV difference between the primary and secondary components is $0.2\kms$ with a  standard deviation of $2\kms$.

\section{Conclusions}

The MEarth team and collaborators are creating a well-studied sample of nearby M dwarfs which will be the basis for future studies investigating their fundamental properties, their evolution, and the exoplanets orbiting them. The data set being assembled is diverse, with photometric rotation periods, parallaxes, and optical spectra. In this work, we presented metallicities, NIR spectral types and radial velocities for a fifth of MEarth M dwarfs.

We created a NIR spectral typing routine, determined by-eye spectral types and presented spectral standards for M1V-M8/9V dwarfs. We related NIR spectral type to PMSU spectral type, finding the conversion to be metallicity-sensitive. We calibrated a new spectroscopic distance relation using NIR spectral type or $\hind$, which can be used to estimate distances to 14\%.

We used M dwarfs in CPM pairs with an F, G or K star of known metallicity to calibrate an empirical metallicity relation.  We validated the physical association of these pairs using proper motions, radial velocities and distances (making use of our RV measurements and spectroscopic distance estimates for the secondaries). We explored the NIR for combinations of EWs that effectively trace stellar metallicity, and found that the EW of the $\na$ line at $2.2\micron$ is sufficient. Our metallicity calibration has a standard deviation of $0.12\dex$ and $R_{ap}=0.78$. It is calibrated using \ncals M dwarfs with NIR spectral types from M1V to M5V and $-0.6<\feh<0.3$, and can be extrapolated to $\feh=-1.0\dex$. We found no evidence that the calibration breaks down for M dwarfs as late as M7V.

Using our $\ewna$ measurements of \nms M dwarfs and the $J-H$, $H-K_S$ color-color diagram, we calibrated a relationship between an M dwarf's distance from the Bessell \& Brett main sequence and its sodium equivalent width. It is valid from $2.5<\ewna(\mathrm{\AA})<7.5$. The standard deviation of our fit is $2$\AA\ and has an $\Rap$ value of $0.92$. Metal-rich M dwarfs can be selected by taking those M dwarfs whose $J-K_S$ colors are redder than the \citet[]{Bessell1988} M dwarf track in the $J-H$, $H-K_S$ color-color diagram.

We developed a method to wavelength calibrate \spex\ M dwarf spectra using telluric features present in the data, and we measured absolute radial velocities for the stars in our sample at a precision of $4.4\kms$. We used synthetic spectra, M dwarfs with precise radial velocities from \citet{Chubak2012} and M dwarf-M dwarf binaries to validate our method. Because telluric absorption features are strong in even short exposure data, our method for determining the absolute wavelength calibration requires no information beyond the science spectrum itself. This opens up the possibility of measuring radial velocities for stars with an extant moderate resolution NIR spectrum.

Our measurements, including NIR spectral types, EWs, radial velocities, and spectroscopic distance estimates are presented in Table A1. We also include distances estimated from parallaxes, and radial velocities from PMSU.
To facilitate joint use of our datasets, we reproduce spectral measurements for M dwarfs observed by R12 in Table A2, with EWs modified to account for differences between their TripleSpec and our IRTF measurements and $\feh$ inferred using our calibration; we also include PMSU spectral types and RVs, and the parallaxes reported in R12. 

In future work, will continue to explore the use of the NIR as a diagnostic of intrinsic stellar properties, investigating how metallicity relates to rotation period, tracers of magnetic activity, and galactic kinematics. 

\acknowledgements

ERN is supported a National Science Foundation Graduate Research Fellowship. The MEarth team gratefully acknowledges funding from the David and Lucile Packard Fellowship for Science and Engineering (awarded to DC). We thank S. Dhital, A. Dupree, M. Holman, and A. West for helpful conversations. This material is based upon work supported by the National Science Foundation under grant number AST-0807690 and AST-1109468. Based on observations at the Infrared Telescope Facility, which is operated by the University of Hawaii under Cooperative Agreement no. NNX-08AE38A with the National Aeronautics and Space Administration, Science Mission Directorate, Planetary Astronomy Program. This research has made extensive use of data products from the Two Micron All Sky Survey, which is a joint project of the University of Massachusetts and the Infrared Processing and Analysis Center / California Institute of Technology, funded by NASA and the NSF, NASAÕs Astrophysics Data System (ADS), and the SIMBAD database, operated at CDS, Strasbourg, France.

\appendix

Tables A1 (All M dwarfs from our rotation and nearby samples and the potential calibrators) and A2 (M dwarfs observed by R12) are available online and in the refereed version of this article. These tables contain positions, proper motions, spectral measurements, measured radial velocities and those from the literature, estimated distances and those from the literature, and inferred $\feh$.


\end{document}